\documentclass[longauth]{aa_gaia} 

\usepackage{amsmath}
\usepackage{lscape}
\usepackage{capt-of}
\usepackage{graphicx}
\usepackage{placeins}
\usepackage[utf8]{inputenc}
\usepackage{filecontents}
\usepackage{tablefootnote}
\usepackage{txfonts}
\usepackage{natbib}
\usepackage{arydshln}
\usepackage{color}

\newcommand{\gaia}{\textit{Gaia \,}}

\def\gaia{\textit{Gaia}\xspace}

\def\gmag{$G$\xspace}
\def\gbp{$G_{\rm BP}$\xspace}
\def\grp{$G_{\rm RP}$\xspace}
\newcommand{\bpminrp}{\ensuremath{G_\mathrm{BP}-G_\mathrm{RP}}\xspace}

\begin{document} 

\titlerunning{\gaia DR2:\hbox{}  CaMD of Variable Stars}
\title{\gaia Data Release 2}
\subtitle{Variable stars in the colour-absolute magnitude diagram} 

\author{
{\it Gaia} Collaboration
\and L.        ~Eyer                          \inst{\ref{inst:0001}}
\and L.        ~Rimoldini                     \inst{\ref{inst:0002}}
\and M.        ~Audard                        \inst{\ref{inst:0001}}
\and R.I.      ~Anderson                      \inst{\ref{inst:0004},\ref{inst:0001}}
\and K.        ~Nienartowicz                  \inst{\ref{inst:0002}}
\and F.        ~Glass                         \inst{\ref{inst:0001}}
\and O.        ~Marchal                       \inst{\ref{inst:0008}}
\and M.        ~Grenon                        \inst{\ref{inst:0001}}
\and N.        ~Mowlavi                       \inst{\ref{inst:0001}}
\and B.        ~Holl                          \inst{\ref{inst:0001}}
\and G.        ~Clementini                    \inst{\ref{inst:0012}}
\and C.        ~Aerts                         \inst{\ref{inst:0013},\ref{inst:0014}}
\and T.        ~Mazeh                         \inst{\ref{inst:0015}}
\and D.W.      ~Evans                         \inst{\ref{inst:0016}}
\and L.        ~Szabados                      \inst{\ref{inst:0017}}
\and A.G.A.    ~Brown                         \inst{\ref{inst:0018}}
\and A.        ~Vallenari                     \inst{\ref{inst:0019}}
\and T.        ~Prusti                        \inst{\ref{inst:0020}}
\and J.H.J.    ~de Bruijne                    \inst{\ref{inst:0020}}
\and C.        ~Babusiaux                     \inst{\ref{inst:0008},\ref{inst:0023}}
\and C.A.L.    ~Bailer-Jones                  \inst{\ref{inst:0024}}
\and M.        ~Biermann                      \inst{\ref{inst:0025}}
\and F.        ~Jansen                        \inst{\ref{inst:0026}}
\and C.        ~Jordi                         \inst{\ref{inst:0027}}
\and S.A.      ~Klioner                       \inst{\ref{inst:0028}}
\and U.        ~Lammers                       \inst{\ref{inst:0029}}
\and L.        ~Lindegren                     \inst{\ref{inst:0030}}
\and X.        ~Luri                          \inst{\ref{inst:0027}}
\and F.        ~Mignard                       \inst{\ref{inst:0032}}
\and C.        ~Panem                         \inst{\ref{inst:0033}}
\and D.        ~Pourbaix                      \inst{\ref{inst:0034},\ref{inst:0035}}
\and S.        ~Randich                       \inst{\ref{inst:0036}}
\and P.        ~Sartoretti                    \inst{\ref{inst:0008}}
\and H.I.      ~Siddiqui                      \inst{\ref{inst:0038}}
\and C.        ~Soubiran                      \inst{\ref{inst:0039}}
\and F.        ~van Leeuwen                   \inst{\ref{inst:0016}}
\and N.A.      ~Walton                        \inst{\ref{inst:0016}}
\and F.        ~Arenou                        \inst{\ref{inst:0008}}
\and U.        ~Bastian                       \inst{\ref{inst:0025}}
\and M.        ~Cropper                       \inst{\ref{inst:0044}}
\and R.        ~Drimmel                       \inst{\ref{inst:0045}}
\and D.        ~Katz                          \inst{\ref{inst:0008}}
\and M.G.      ~Lattanzi                      \inst{\ref{inst:0045}}
\and J.        ~Bakker                        \inst{\ref{inst:0029}}
\and C.        ~Cacciari                      \inst{\ref{inst:0012}}
\and J.        ~Casta\~{n}eda                 \inst{\ref{inst:0027}}
\and L.        ~Chaoul                        \inst{\ref{inst:0033}}
\and N.        ~Cheek                         \inst{\ref{inst:0052}}
\and F.        ~De Angeli                     \inst{\ref{inst:0016}}
\and C.        ~Fabricius                     \inst{\ref{inst:0027}}
\and R.        ~Guerra                        \inst{\ref{inst:0029}}
\and E.        ~Masana                        \inst{\ref{inst:0027}}
\and R.        ~Messineo                      \inst{\ref{inst:0057}}
\and P.        ~Panuzzo                       \inst{\ref{inst:0008}}
\and J.        ~Portell                       \inst{\ref{inst:0027}}
\and M.        ~Riello                        \inst{\ref{inst:0016}}
\and G.M.      ~Seabroke                      \inst{\ref{inst:0044}}
\and P.        ~Tanga                         \inst{\ref{inst:0032}}
\and F.        ~Th\'{e}venin                  \inst{\ref{inst:0032}}
\and G.        ~Gracia-Abril                  \inst{\ref{inst:0064},\ref{inst:0025}}
\and G.        ~Comoretto                     \inst{\ref{inst:0038}}
\and M.        ~Garcia-Reinaldos              \inst{\ref{inst:0029}}
\and D.        ~Teyssier                      \inst{\ref{inst:0038}}
\and M.        ~Altmann                       \inst{\ref{inst:0025},\ref{inst:0070}}
\and R.        ~Andrae                        \inst{\ref{inst:0024}}
\and I.        ~Bellas-Velidis                \inst{\ref{inst:0072}}
\and K.        ~Benson                        \inst{\ref{inst:0044}}
\and J.        ~Berthier                      \inst{\ref{inst:0074}}
\and R.        ~Blomme                        \inst{\ref{inst:0075}}
\and P.        ~Burgess                       \inst{\ref{inst:0016}}
\and G.        ~Busso                         \inst{\ref{inst:0016}}
\and B.        ~Carry                         \inst{\ref{inst:0032},\ref{inst:0074}}
\and A.        ~Cellino                       \inst{\ref{inst:0045}}
\and M.        ~Clotet                        \inst{\ref{inst:0027}}
\and O.        ~Creevey                       \inst{\ref{inst:0032}}
\and M.        ~Davidson                      \inst{\ref{inst:0083}}
\and J.        ~De Ridder                     \inst{\ref{inst:0013}}
\and L.        ~Delchambre                    \inst{\ref{inst:0085}}
\and A.        ~Dell'Oro                      \inst{\ref{inst:0036}}
\and C.        ~Ducourant                     \inst{\ref{inst:0039}}
\and J.        ~Fern\'{a}ndez-Hern\'{a}ndez   \inst{\ref{inst:0088}}
\and M.        ~Fouesneau                     \inst{\ref{inst:0024}}
\and Y.        ~Fr\'{e}mat                    \inst{\ref{inst:0075}}
\and L.        ~Galluccio                     \inst{\ref{inst:0032}}
\and M.        ~Garc\'{i}a-Torres             \inst{\ref{inst:0092}}
\and J.        ~Gonz\'{a}lez-N\'{u}\~{n}ez    \inst{\ref{inst:0052},\ref{inst:0094}}
\and J.J.      ~Gonz\'{a}lez-Vidal            \inst{\ref{inst:0027}}
\and E.        ~Gosset                        \inst{\ref{inst:0085},\ref{inst:0035}}
\and L.P.      ~Guy                           \inst{\ref{inst:0002},\ref{inst:0099}}
\and J.-L.     ~Halbwachs                     \inst{\ref{inst:0100}}
\and N.C.      ~Hambly                        \inst{\ref{inst:0083}}
\and D.L.      ~Harrison                      \inst{\ref{inst:0016},\ref{inst:0103}}
\and J.        ~Hern\'{a}ndez                 \inst{\ref{inst:0029}}
\and D.        ~Hestroffer                    \inst{\ref{inst:0074}}
\and S.T.      ~Hodgkin                       \inst{\ref{inst:0016}}
\and A.        ~Hutton                        \inst{\ref{inst:0107}}
\and G.        ~Jasniewicz                    \inst{\ref{inst:0108}}
\and A.        ~Jean-Antoine-Piccolo          \inst{\ref{inst:0033}}
\and S.        ~Jordan                        \inst{\ref{inst:0025}}
\and A.J.      ~Korn                          \inst{\ref{inst:0111}}
\and A.        ~Krone-Martins                 \inst{\ref{inst:0112}}
\and A.C.      ~Lanzafame                     \inst{\ref{inst:0113},\ref{inst:0114}}
\and T.        ~Lebzelter                     \inst{\ref{inst:0115}}
\and W.        ~L\"{ o}ffler                  \inst{\ref{inst:0025}}
\and M.        ~Manteiga                      \inst{\ref{inst:0117},\ref{inst:0118}}
\and P.M.      ~Marrese                       \inst{\ref{inst:0119},\ref{inst:0120}}
\and J.M.      ~Mart\'{i}n-Fleitas            \inst{\ref{inst:0107}}
\and A.        ~Moitinho                      \inst{\ref{inst:0112}}
\and A.        ~Mora                          \inst{\ref{inst:0107}}
\and K.        ~Muinonen                      \inst{\ref{inst:0124},\ref{inst:0125}}
\and J.        ~Osinde                        \inst{\ref{inst:0126}}
\and E.        ~Pancino                       \inst{\ref{inst:0036},\ref{inst:0120}}
\and T.        ~Pauwels                       \inst{\ref{inst:0075}}
\and J.-M.     ~Petit                         \inst{\ref{inst:0130}}
\and A.        ~Recio-Blanco                  \inst{\ref{inst:0032}}
\and P.J.      ~Richards                      \inst{\ref{inst:0132}}
\and A.C.      ~Robin                         \inst{\ref{inst:0130}}
\and L.M.      ~Sarro                         \inst{\ref{inst:0134}}
\and C.        ~Siopis                        \inst{\ref{inst:0034}}
\and M.        ~Smith                         \inst{\ref{inst:0044}}
\and A.        ~Sozzetti                      \inst{\ref{inst:0045}}
\and M.        ~S\"{ u}veges                  \inst{\ref{inst:0024}}
\and J.        ~Torra                         \inst{\ref{inst:0027}}
\and W.        ~van Reeven                    \inst{\ref{inst:0107}}
\and U.        ~Abbas                         \inst{\ref{inst:0045}}
\and A.        ~Abreu Aramburu                \inst{\ref{inst:0142}}
\and S.        ~Accart                        \inst{\ref{inst:0143}}
\and G.        ~Altavilla                     \inst{\ref{inst:0119},\ref{inst:0120},\ref{inst:0012}}
\and M.A.      ~\'{A}lvarez                   \inst{\ref{inst:0117}}
\and R.        ~Alvarez                       \inst{\ref{inst:0029}}
\and J.        ~Alves                         \inst{\ref{inst:0115}}
\and A.H.      ~Andrei                        \inst{\ref{inst:0150},\ref{inst:0151},\ref{inst:0070}}
\and E.        ~Anglada Varela                \inst{\ref{inst:0088}}
\and E.        ~Antiche                       \inst{\ref{inst:0027}}
\and T.        ~Antoja                        \inst{\ref{inst:0020},\ref{inst:0027}}
\and B.        ~Arcay                         \inst{\ref{inst:0117}}
\and T.L.      ~Astraatmadja                  \inst{\ref{inst:0024},\ref{inst:0159}}
\and N.        ~Bach                          \inst{\ref{inst:0107}}
\and S.G.      ~Baker                         \inst{\ref{inst:0044}}
\and L.        ~Balaguer-N\'{u}\~{n}ez        \inst{\ref{inst:0027}}
\and P.        ~Balm                          \inst{\ref{inst:0038}}
\and C.        ~Barache                       \inst{\ref{inst:0070}}
\and C.        ~Barata                        \inst{\ref{inst:0112}}
\and D.        ~Barbato                       \inst{\ref{inst:0166},\ref{inst:0045}}
\and F.        ~Barblan                       \inst{\ref{inst:0001}}
\and P.S.      ~Barklem                       \inst{\ref{inst:0111}}
\and D.        ~Barrado                       \inst{\ref{inst:0170}}
\and M.        ~Barros                        \inst{\ref{inst:0112}}
\and M.A.      ~Barstow                       \inst{\ref{inst:0172}}
\and S.        ~Bartholom\'{e} Mu\~{n}oz      \inst{\ref{inst:0027}}
\and J.-L.     ~Bassilana                     \inst{\ref{inst:0143}}
\and U.        ~Becciani                      \inst{\ref{inst:0114}}
\and M.        ~Bellazzini                    \inst{\ref{inst:0012}}
\and A.        ~Berihuete                     \inst{\ref{inst:0177}}
\and S.        ~Bertone                       \inst{\ref{inst:0045},\ref{inst:0070},\ref{inst:0180}}
\and L.        ~Bianchi                       \inst{\ref{inst:0181}}
\and O.        ~Bienaym\'{e}                  \inst{\ref{inst:0100}}
\and S.        ~Blanco-Cuaresma               \inst{\ref{inst:0001},\ref{inst:0039},\ref{inst:0185}}
\and T.        ~Boch                          \inst{\ref{inst:0100}}
\and C.        ~Boeche                        \inst{\ref{inst:0019}}
\and A.        ~Bombrun                       \inst{\ref{inst:0188}}
\and R.        ~Borrachero                    \inst{\ref{inst:0027}}
\and D.        ~Bossini                       \inst{\ref{inst:0019}}
\and S.        ~Bouquillon                    \inst{\ref{inst:0070}}
\and G.        ~Bourda                        \inst{\ref{inst:0039}}
\and A.        ~Bragaglia                     \inst{\ref{inst:0012}}
\and L.        ~Bramante                      \inst{\ref{inst:0057}}
\and M.A.      ~Breddels                      \inst{\ref{inst:0195}}
\and A.        ~Bressan                       \inst{\ref{inst:0196}}
\and N.        ~Brouillet                     \inst{\ref{inst:0039}}
\and T.        ~Br\"{ u}semeister             \inst{\ref{inst:0025}}
\and E.        ~Brugaletta                    \inst{\ref{inst:0114}}
\and B.        ~Bucciarelli                   \inst{\ref{inst:0045}}
\and A.        ~Burlacu                       \inst{\ref{inst:0033}}
\and D.        ~Busonero                      \inst{\ref{inst:0045}}
\and A.G.      ~Butkevich                     \inst{\ref{inst:0028}}
\and R.        ~Buzzi                         \inst{\ref{inst:0045}}
\and E.        ~Caffau                        \inst{\ref{inst:0008}}
\and R.        ~Cancelliere                   \inst{\ref{inst:0206}}
\and G.        ~Cannizzaro                    \inst{\ref{inst:0207},\ref{inst:0014}}
\and T.        ~Cantat-Gaudin                 \inst{\ref{inst:0019},\ref{inst:0027}}
\and R.        ~Carballo                      \inst{\ref{inst:0211}}
\and T.        ~Carlucci                      \inst{\ref{inst:0070}}
\and J.M.      ~Carrasco                      \inst{\ref{inst:0027}}
\and L.        ~Casamiquela                   \inst{\ref{inst:0027}}
\and M.        ~Castellani                    \inst{\ref{inst:0119}}
\and A.        ~Castro-Ginard                 \inst{\ref{inst:0027}}
\and P.        ~Charlot                       \inst{\ref{inst:0039}}
\and L.        ~Chemin                        \inst{\ref{inst:0218}}
\and A.        ~Chiavassa                     \inst{\ref{inst:0032}}
\and G.        ~Cocozza                       \inst{\ref{inst:0012}}
\and G.        ~Costigan                      \inst{\ref{inst:0018}}
\and S.        ~Cowell                        \inst{\ref{inst:0016}}
\and F.        ~Crifo                         \inst{\ref{inst:0008}}
\and M.        ~Crosta                        \inst{\ref{inst:0045}}
\and C.        ~Crowley                       \inst{\ref{inst:0188}}
\and J.        ~Cuypers$^\dagger$             \inst{\ref{inst:0075}}
\and C.        ~Dafonte                       \inst{\ref{inst:0117}}
\and Y.        ~Damerdji                      \inst{\ref{inst:0085},\ref{inst:0229}}
\and A.        ~Dapergolas                    \inst{\ref{inst:0072}}
\and P.        ~David                         \inst{\ref{inst:0074}}
\and M.        ~David                         \inst{\ref{inst:0232}}
\and P.        ~de Laverny                    \inst{\ref{inst:0032}}
\and F.        ~De Luise                      \inst{\ref{inst:0234}}
\and R.        ~De March                      \inst{\ref{inst:0057}}
\and D.        ~de Martino                    \inst{\ref{inst:0236}}
\and R.        ~de Souza                      \inst{\ref{inst:0237}}
\and A.        ~de Torres                     \inst{\ref{inst:0188}}
\and J.        ~Debosscher                    \inst{\ref{inst:0013}}
\and E.        ~del Pozo                      \inst{\ref{inst:0107}}
\and M.        ~Delbo                         \inst{\ref{inst:0032}}
\and A.        ~Delgado                       \inst{\ref{inst:0016}}
\and H.E.      ~Delgado                       \inst{\ref{inst:0134}}
\and S.        ~Diakite                       \inst{\ref{inst:0130}}
\and C.        ~Diener                        \inst{\ref{inst:0016}}
\and E.        ~Distefano                     \inst{\ref{inst:0114}}
\and C.        ~Dolding                       \inst{\ref{inst:0044}}
\and P.        ~Drazinos                      \inst{\ref{inst:0248}}
\and J.        ~Dur\'{a}n                     \inst{\ref{inst:0126}}
\and B.        ~Edvardsson                    \inst{\ref{inst:0111}}
\and H.        ~Enke                          \inst{\ref{inst:0251}}
\and K.        ~Eriksson                      \inst{\ref{inst:0111}}
\and P.        ~Esquej                        \inst{\ref{inst:0253}}
\and G.        ~Eynard Bontemps               \inst{\ref{inst:0033}}
\and C.        ~Fabre                         \inst{\ref{inst:0255}}
\and M.        ~Fabrizio                      \inst{\ref{inst:0119},\ref{inst:0120}}
\and S.        ~Faigler                       \inst{\ref{inst:0015}}
\and A.J.      ~Falc\~{a}o                    \inst{\ref{inst:0259}}
\and M.        ~Farr\`{a}s Casas              \inst{\ref{inst:0027}}
\and L.        ~Federici                      \inst{\ref{inst:0012}}
\and G.        ~Fedorets                      \inst{\ref{inst:0124}}
\and P.        ~Fernique                      \inst{\ref{inst:0100}}
\and F.        ~Figueras                      \inst{\ref{inst:0027}}
\and F.        ~Filippi                       \inst{\ref{inst:0057}}
\and K.        ~Findeisen                     \inst{\ref{inst:0008}}
\and A.        ~Fonti                         \inst{\ref{inst:0057}}
\and E.        ~Fraile                        \inst{\ref{inst:0253}}
\and M.        ~Fraser                        \inst{\ref{inst:0016},\ref{inst:0270}}
\and B.        ~Fr\'{e}zouls                  \inst{\ref{inst:0033}}
\and M.        ~Gai                           \inst{\ref{inst:0045}}
\and S.        ~Galleti                       \inst{\ref{inst:0012}}
\and D.        ~Garabato                      \inst{\ref{inst:0117}}
\and F.        ~Garc\'{i}a-Sedano             \inst{\ref{inst:0134}}
\and A.        ~Garofalo                      \inst{\ref{inst:0276},\ref{inst:0012}}
\and N.        ~Garralda                      \inst{\ref{inst:0027}}
\and A.        ~Gavel                         \inst{\ref{inst:0111}}
\and P.        ~Gavras                        \inst{\ref{inst:0008},\ref{inst:0072},\ref{inst:0248}}
\and J.        ~Gerssen                       \inst{\ref{inst:0251}}
\and R.        ~Geyer                         \inst{\ref{inst:0028}}
\and P.        ~Giacobbe                      \inst{\ref{inst:0045}}
\and G.        ~Gilmore                       \inst{\ref{inst:0016}}
\and S.        ~Girona                        \inst{\ref{inst:0287}}
\and G.        ~Giuffrida                     \inst{\ref{inst:0120},\ref{inst:0119}}
\and M.        ~Gomes                         \inst{\ref{inst:0112}}
\and M.        ~Granvik                       \inst{\ref{inst:0124},\ref{inst:0292}}
\and A.        ~Gueguen                       \inst{\ref{inst:0008},\ref{inst:0294}}
\and A.        ~Guerrier                      \inst{\ref{inst:0143}}
\and J.        ~Guiraud                       \inst{\ref{inst:0033}}
\and R.        ~Guti\'{e}rrez-S\'{a}nchez     \inst{\ref{inst:0038}}
\and R.        ~Haigron                       \inst{\ref{inst:0008}}
\and D.        ~Hatzidimitriou                \inst{\ref{inst:0248},\ref{inst:0072}}
\and M.        ~Hauser                        \inst{\ref{inst:0025},\ref{inst:0024}}
\and M.        ~Haywood                       \inst{\ref{inst:0008}}
\and U.        ~Heiter                        \inst{\ref{inst:0111}}
\and A.        ~Helmi                         \inst{\ref{inst:0195}}
\and J.        ~Heu                           \inst{\ref{inst:0008}}
\and T.        ~Hilger                        \inst{\ref{inst:0028}}
\and D.        ~Hobbs                         \inst{\ref{inst:0030}}
\and W.        ~Hofmann                       \inst{\ref{inst:0025}}
\and G.        ~Holland                       \inst{\ref{inst:0016}}
\and H.E.      ~Huckle                        \inst{\ref{inst:0044}}
\and A.        ~Hypki                         \inst{\ref{inst:0018},\ref{inst:0313}}
\and V.        ~Icardi                        \inst{\ref{inst:0057}}
\and K.        ~Jan{\ss}en                    \inst{\ref{inst:0251}}
\and G.        ~Jevardat de Fombelle          \inst{\ref{inst:0002}}
\and P.G.      ~Jonker                        \inst{\ref{inst:0207},\ref{inst:0014}}
\and \'{A}.L.  ~Juh\'{a}sz                    \inst{\ref{inst:0017},\ref{inst:0320}}
\and F.        ~Julbe                         \inst{\ref{inst:0027}}
\and A.        ~Karampelas                    \inst{\ref{inst:0248},\ref{inst:0323}}
\and A.        ~Kewley                        \inst{\ref{inst:0016}}
\and J.        ~Klar                          \inst{\ref{inst:0251}}
\and A.        ~Kochoska                      \inst{\ref{inst:0326},\ref{inst:0327}}
\and R.        ~Kohley                        \inst{\ref{inst:0029}}
\and K.        ~Kolenberg                     \inst{\ref{inst:0329},\ref{inst:0013},\ref{inst:0185}}
\and M.        ~Kontizas                      \inst{\ref{inst:0248}}
\and E.        ~Kontizas                      \inst{\ref{inst:0072}}
\and S.E.      ~Koposov                       \inst{\ref{inst:0016},\ref{inst:0335}}
\and G.        ~Kordopatis                    \inst{\ref{inst:0032}}
\and Z.        ~Kostrzewa-Rutkowska           \inst{\ref{inst:0207},\ref{inst:0014}}
\and P.        ~Koubsky                       \inst{\ref{inst:0339}}
\and S.        ~Lambert                       \inst{\ref{inst:0070}}
\and A.F.      ~Lanza                         \inst{\ref{inst:0114}}
\and Y.        ~Lasne                         \inst{\ref{inst:0143}}
\and J.-B.     ~Lavigne                       \inst{\ref{inst:0143}}
\and Y.        ~Le Fustec                     \inst{\ref{inst:0344}}
\and C.        ~Le Poncin-Lafitte             \inst{\ref{inst:0070}}
\and Y.        ~Lebreton                      \inst{\ref{inst:0008},\ref{inst:0347}}
\and S.        ~Leccia                        \inst{\ref{inst:0236}}
\and N.        ~Leclerc                       \inst{\ref{inst:0008}}
\and I.        ~Lecoeur-Taibi                 \inst{\ref{inst:0002}}
\and H.        ~Lenhardt                      \inst{\ref{inst:0025}}
\and F.        ~Leroux                        \inst{\ref{inst:0143}}
\and S.        ~Liao                          \inst{\ref{inst:0045},\ref{inst:0354},\ref{inst:0355}}
\and E.        ~Licata                        \inst{\ref{inst:0181}}
\and H.E.P.    ~Lindstr{\o}m                  \inst{\ref{inst:0357},\ref{inst:0358}}
\and T.A.      ~Lister                        \inst{\ref{inst:0359}}
\and E.        ~Livanou                       \inst{\ref{inst:0248}}
\and A.        ~Lobel                         \inst{\ref{inst:0075}}
\and M.        ~L\'{o}pez                     \inst{\ref{inst:0170}}
\and D.        ~Lorenz                        \inst{\ref{inst:0115}}
\and S.        ~Managau                       \inst{\ref{inst:0143}}
\and R.G.      ~Mann                          \inst{\ref{inst:0083}}
\and G.        ~Mantelet                      \inst{\ref{inst:0025}}
\and J.M.      ~Marchant                      \inst{\ref{inst:0367}}
\and M.        ~Marconi                       \inst{\ref{inst:0236}}
\and S.        ~Marinoni                      \inst{\ref{inst:0119},\ref{inst:0120}}
\and G.        ~Marschalk\'{o}                \inst{\ref{inst:0017},\ref{inst:0372}}
\and D.J.      ~Marshall                      \inst{\ref{inst:0373}}
\and M.        ~Martino                       \inst{\ref{inst:0057}}
\and G.        ~Marton                        \inst{\ref{inst:0017}}
\and N.        ~Mary                          \inst{\ref{inst:0143}}
\and D.        ~Massari                       \inst{\ref{inst:0195}}
\and G.        ~Matijevi\v{c}                 \inst{\ref{inst:0251}}
\and P.J.      ~McMillan                      \inst{\ref{inst:0030}}
\and S.        ~Messina                       \inst{\ref{inst:0114}}
\and D.        ~Michalik                      \inst{\ref{inst:0030}}
\and N.R.      ~Millar                        \inst{\ref{inst:0016}}
\and D.        ~Molina                        \inst{\ref{inst:0027}}
\and R.        ~Molinaro                      \inst{\ref{inst:0236}}
\and L.        ~Moln\'{a}r                    \inst{\ref{inst:0017}}
\and P.        ~Montegriffo                   \inst{\ref{inst:0012}}
\and R.        ~Mor                           \inst{\ref{inst:0027}}
\and R.        ~Morbidelli                    \inst{\ref{inst:0045}}
\and T.        ~Morel                         \inst{\ref{inst:0085}}
\and S.        ~Morgenthaler                  \inst{\ref{inst:0390}}
\and D.        ~Morris                        \inst{\ref{inst:0083}}
\and A.F.      ~Mulone                        \inst{\ref{inst:0057}}
\and T.        ~Muraveva                      \inst{\ref{inst:0012}}
\and I.        ~Musella                       \inst{\ref{inst:0236}}
\and G.        ~Nelemans                      \inst{\ref{inst:0014},\ref{inst:0013}}
\and L.        ~Nicastro                      \inst{\ref{inst:0012}}
\and L.        ~Noval                         \inst{\ref{inst:0143}}
\and W.        ~O'Mullane                     \inst{\ref{inst:0029},\ref{inst:0099}}
\and C.        ~Ord\'{e}novic                 \inst{\ref{inst:0032}}
\and D.        ~Ord\'{o}\~{n}ez-Blanco        \inst{\ref{inst:0002}}
\and P.        ~Osborne                       \inst{\ref{inst:0016}}
\and C.        ~Pagani                        \inst{\ref{inst:0172}}
\and I.        ~Pagano                        \inst{\ref{inst:0114}}
\and F.        ~Pailler                       \inst{\ref{inst:0033}}
\and H.        ~Palacin                       \inst{\ref{inst:0143}}
\and L.        ~Palaversa                     \inst{\ref{inst:0016},\ref{inst:0001}}
\and A.        ~Panahi                        \inst{\ref{inst:0015}}
\and M.        ~Pawlak                        \inst{\ref{inst:0411},\ref{inst:0412}}
\and A.M.      ~Piersimoni                    \inst{\ref{inst:0234}}
\and F.-X.     ~Pineau                        \inst{\ref{inst:0100}}
\and E.        ~Plachy                        \inst{\ref{inst:0017}}
\and G.        ~Plum                          \inst{\ref{inst:0008}}
\and E.        ~Poggio                        \inst{\ref{inst:0166},\ref{inst:0045}}
\and E.        ~Poujoulet                     \inst{\ref{inst:0419}}
\and A.        ~Pr\v{s}a                      \inst{\ref{inst:0327}}
\and L.        ~Pulone                        \inst{\ref{inst:0119}}
\and E.        ~Racero                        \inst{\ref{inst:0052}}
\and S.        ~Ragaini                       \inst{\ref{inst:0012}}
\and N.        ~Rambaux                       \inst{\ref{inst:0074}}
\and M.        ~Ramos-Lerate                  \inst{\ref{inst:0425}}
\and S.        ~Regibo                        \inst{\ref{inst:0013}}
\and C.        ~Reyl\'{e}                     \inst{\ref{inst:0130}}
\and F.        ~Riclet                        \inst{\ref{inst:0033}}
\and V.        ~Ripepi                        \inst{\ref{inst:0236}}
\and A.        ~Riva                          \inst{\ref{inst:0045}}
\and A.        ~Rivard                        \inst{\ref{inst:0143}}
\and G.        ~Rixon                         \inst{\ref{inst:0016}}
\and T.        ~Roegiers                      \inst{\ref{inst:0433}}
\and M.        ~Roelens                       \inst{\ref{inst:0001}}
\and M.        ~Romero-G\'{o}mez              \inst{\ref{inst:0027}}
\and N.        ~Rowell                        \inst{\ref{inst:0083}}
\and F.        ~Royer                         \inst{\ref{inst:0008}}
\and L.        ~Ruiz-Dern                     \inst{\ref{inst:0008}}
\and G.        ~Sadowski                      \inst{\ref{inst:0034}}
\and T.        ~Sagrist\`{a} Sell\'{e}s       \inst{\ref{inst:0025}}
\and J.        ~Sahlmann                      \inst{\ref{inst:0029},\ref{inst:0442}}
\and J.        ~Salgado                       \inst{\ref{inst:0443}}
\and E.        ~Salguero                      \inst{\ref{inst:0088}}
\and N.        ~Sanna                         \inst{\ref{inst:0036}}
\and T.        ~Santana-Ros                   \inst{\ref{inst:0313}}
\and M.        ~Sarasso                       \inst{\ref{inst:0045}}
\and H.        ~Savietto                      \inst{\ref{inst:0448}}
\and M.        ~Schultheis                    \inst{\ref{inst:0032}}
\and E.        ~Sciacca                       \inst{\ref{inst:0114}}
\and M.        ~Segol                         \inst{\ref{inst:0451}}
\and J.C.      ~Segovia                       \inst{\ref{inst:0052}}
\and D.        ~S\'{e}gransan                 \inst{\ref{inst:0001}}
\and I-C.      ~Shih                          \inst{\ref{inst:0008}}
\and L.        ~Siltala                       \inst{\ref{inst:0124},\ref{inst:0456}}
\and A.F.      ~Silva                         \inst{\ref{inst:0112}}
\and R.L.      ~Smart                         \inst{\ref{inst:0045}}
\and K.W.      ~Smith                         \inst{\ref{inst:0024}}
\and E.        ~Solano                        \inst{\ref{inst:0170},\ref{inst:0461}}
\and F.        ~Solitro                       \inst{\ref{inst:0057}}
\and R.        ~Sordo                         \inst{\ref{inst:0019}}
\and S.        ~Soria Nieto                   \inst{\ref{inst:0027}}
\and J.        ~Souchay                       \inst{\ref{inst:0070}}
\and A.        ~Spagna                        \inst{\ref{inst:0045}}
\and F.        ~Spoto                         \inst{\ref{inst:0032},\ref{inst:0074}}
\and U.        ~Stampa                        \inst{\ref{inst:0025}}
\and I.A.      ~Steele                        \inst{\ref{inst:0367}}
\and H.        ~Steidelm\"{ u}ller            \inst{\ref{inst:0028}}
\and C.A.      ~Stephenson                    \inst{\ref{inst:0038}}
\and H.        ~Stoev                         \inst{\ref{inst:0473}}
\and F.F.      ~Suess                         \inst{\ref{inst:0016}}
\and J.        ~Surdej                        \inst{\ref{inst:0085}}
\and E.        ~Szegedi-Elek                  \inst{\ref{inst:0017}}
\and D.        ~Tapiador                      \inst{\ref{inst:0477},\ref{inst:0478}}
\and F.        ~Taris                         \inst{\ref{inst:0070}}
\and G.        ~Tauran                        \inst{\ref{inst:0143}}
\and M.B.      ~Taylor                        \inst{\ref{inst:0481}}
\and R.        ~Teixeira                      \inst{\ref{inst:0237}}
\and D.        ~Terrett                       \inst{\ref{inst:0132}}
\and P.        ~Teyssandier                   \inst{\ref{inst:0070}}
\and W.        ~Thuillot                      \inst{\ref{inst:0074}}
\and A.        ~Titarenko                     \inst{\ref{inst:0032}}
\and F.        ~Torra Clotet                  \inst{\ref{inst:0487}}
\and C.        ~Turon                         \inst{\ref{inst:0008}}
\and A.        ~Ulla                          \inst{\ref{inst:0489}}
\and E.        ~Utrilla                       \inst{\ref{inst:0107}}
\and S.        ~Uzzi                          \inst{\ref{inst:0057}}
\and M.        ~Vaillant                      \inst{\ref{inst:0143}}
\and G.        ~Valentini                     \inst{\ref{inst:0234}}
\and V.        ~Valette                       \inst{\ref{inst:0033}}
\and A.        ~van Elteren                   \inst{\ref{inst:0018}}
\and E.        ~Van Hemelryck                 \inst{\ref{inst:0075}}
\and M.        ~van Leeuwen                   \inst{\ref{inst:0016}}
\and M.        ~Vaschetto                     \inst{\ref{inst:0057}}
\and A.        ~Vecchiato                     \inst{\ref{inst:0045}}
\and J.        ~Veljanoski                    \inst{\ref{inst:0195}}
\and Y.        ~Viala                         \inst{\ref{inst:0008}}
\and D.        ~Vicente                       \inst{\ref{inst:0287}}
\and S.        ~Vogt                          \inst{\ref{inst:0433}}
\and C.        ~von Essen                     \inst{\ref{inst:0504}}
\and H.        ~Voss                          \inst{\ref{inst:0027}}
\and V.        ~Votruba                       \inst{\ref{inst:0339}}
\and S.        ~Voutsinas                     \inst{\ref{inst:0083}}
\and G.        ~Walmsley                      \inst{\ref{inst:0033}}
\and M.        ~Weiler                        \inst{\ref{inst:0027}}
\and O.        ~Wertz                         \inst{\ref{inst:0510}}
\and T.        ~Wevers                        \inst{\ref{inst:0016},\ref{inst:0014}}
\and \L{}.     ~Wyrzykowski                   \inst{\ref{inst:0016},\ref{inst:0411}}
\and A.        ~Yoldas                        \inst{\ref{inst:0016}}
\and M.        ~\v{Z}erjal                    \inst{\ref{inst:0326},\ref{inst:0517}}
\and H.        ~Ziaeepour                     \inst{\ref{inst:0130}}
\and J.        ~Zorec                         \inst{\ref{inst:0519}}
\and S.        ~Zschocke                      \inst{\ref{inst:0028}}
\and S.        ~Zucker                        \inst{\ref{inst:0521}}
\and C.        ~Zurbach                       \inst{\ref{inst:0108}}
\and T.        ~Zwitter                       \inst{\ref{inst:0326}}
}
\institute{
     Department of Astronomy, University of Geneva, Chemin des Maillettes 51, 1290 Versoix, Switzerland\relax                                                                                                \label{inst:0001}
\and Department of Astronomy, University of Geneva, Chemin d'Ecogia 16, 1290 Versoix, Switzerland\relax                                                                                                      \label{inst:0002}
\and European Southern Observatory, Karl-Schwarzschild-Str. 2, 85748 Garching, Germany\relax                                                                                                                 \label{inst:0004}
\and GEPI, Observatoire de Paris, Universit\'{e} PSL, CNRS, 5 Place Jules Janssen, 92190 Meudon, France\relax                                                                                                \label{inst:0008}
\and INAF - Osservatorio di Astrofisica e Scienza dello Spazio di Bologna, via Piero Gobetti 93/3, 40129 Bologna, Italy\relax                                                                                \label{inst:0012}
\and Instituut voor Sterrenkunde, KU Leuven, Celestijnenlaan 200D, 3001 Leuven, Belgium\relax                                                                                                                \label{inst:0013}
\and Department of Astrophysics/IMAPP, Radboud University, P.O.Box 9010, 6500 GL Nijmegen, The Netherlands\relax                                                                                             \label{inst:0014}
\and School of Physics and Astronomy, Tel Aviv University, Tel Aviv 6997801, Israel\relax                                                                                                                    \label{inst:0015}
\and Institute of Astronomy, University of Cambridge, Madingley Road, Cambridge CB3 0HA, United Kingdom\relax                                                                                                \label{inst:0016}
\and Konkoly Observatory, Research Centre for Astronomy and Earth Sciences, Hungarian Academy of Sciences, Konkoly Thege Mikl\'{o}s \'{u}t 15-17, 1121 Budapest, Hungary\relax                               \label{inst:0017}
\and Leiden Observatory, Leiden University, Niels Bohrweg 2, 2333 CA Leiden, The Netherlands\relax                                                                                                           \label{inst:0018}
\and INAF - Osservatorio astronomico di Padova, Vicolo Osservatorio 5, 35122 Padova, Italy\relax                                                                                                             \label{inst:0019}
\and Science Support Office, Directorate of Science, European Space Research and Technology Centre (ESA/ESTEC), Keplerlaan 1, 2201AZ, Noordwijk, The Netherlands\relax                                       \label{inst:0020}
\and Univ. Grenoble Alpes, CNRS, IPAG, 38000 Grenoble, France\relax                                                                                                                                          \label{inst:0023}
\and Max Planck Institute for Astronomy, K\"{ o}nigstuhl 17, 69117 Heidelberg, Germany\relax                                                                                                                 \label{inst:0024}
\and Astronomisches Rechen-Institut, Zentrum f\"{ u}r Astronomie der Universit\"{ a}t Heidelberg, M\"{ o}nchhofstr. 12-14, 69120 Heidelberg, Germany\relax                                                   \label{inst:0025}
\and Mission Operations Division, Operations Department, Directorate of Science, European Space Research and Technology Centre (ESA/ESTEC), Keplerlaan 1, 2201 AZ, Noordwijk, The Netherlands\relax          \label{inst:0026}
\and Institut de Ci\`{e}ncies del Cosmos, Universitat  de  Barcelona  (IEEC-UB), Mart\'{i} i  Franqu\`{e}s  1, 08028 Barcelona, Spain\relax                                                                  \label{inst:0027}
\and Lohrmann Observatory, Technische Universit\"{ a}t Dresden, Mommsenstra{\ss}e 13, 01062 Dresden, Germany\relax                                                                                           \label{inst:0028}
\and European Space Astronomy Centre (ESA/ESAC), Camino bajo del Castillo, s/n, Urbanizacion Villafranca del Castillo, Villanueva de la Ca\~{n}ada, 28692 Madrid, Spain\relax                                \label{inst:0029}
\and Lund Observatory, Department of Astronomy and Theoretical Physics, Lund University, Box 43, 22100 Lund, Sweden\relax                                                                                    \label{inst:0030}
\and Universit\'{e} C\^{o}te d'Azur, Observatoire de la C\^{o}te d'Azur, CNRS, Laboratoire Lagrange, Bd de l'Observatoire, CS 34229, 06304 Nice Cedex 4, France\relax                                        \label{inst:0032}
\and CNES Centre Spatial de Toulouse, 18 avenue Edouard Belin, 31401 Toulouse Cedex 9, France\relax                                                                                                          \label{inst:0033}
\and Institut d'Astronomie et d'Astrophysique, Universit\'{e} Libre de Bruxelles CP 226, Boulevard du Triomphe, 1050 Brussels, Belgium\relax                                                                 \label{inst:0034}
\and F.R.S.-FNRS, Rue d'Egmont 5, 1000 Brussels, Belgium\relax                                                                                                                                               \label{inst:0035}
\and INAF - Osservatorio Astrofisico di Arcetri, Largo Enrico Fermi 5, 50125 Firenze, Italy\relax                                                                                                            \label{inst:0036}
\and Telespazio Vega UK Ltd for ESA/ESAC, Camino bajo del Castillo, s/n, Urbanizacion Villafranca del Castillo, Villanueva de la Ca\~{n}ada, 28692 Madrid, Spain\relax                                       \label{inst:0038}
\and Laboratoire d'astrophysique de Bordeaux, Univ. Bordeaux, CNRS, B18N, all{\'e}e Geoffroy Saint-Hilaire, 33615 Pessac, France\relax                                                                       \label{inst:0039}
\and Mullard Space Science Laboratory, University College London, Holmbury St Mary, Dorking, Surrey RH5 6NT, United Kingdom\relax                                                                            \label{inst:0044}
\and INAF - Osservatorio Astrofisico di Torino, via Osservatorio 20, 10025 Pino Torinese (TO), Italy\relax                                                                                                   \label{inst:0045}
\and Serco Gesti\'{o}n de Negocios for ESA/ESAC, Camino bajo del Castillo, s/n, Urbanizacion Villafranca del Castillo, Villanueva de la Ca\~{n}ada, 28692 Madrid, Spain\relax                                \label{inst:0052}
\and ALTEC S.p.a, Corso Marche, 79,10146 Torino, Italy\relax                                                                                                                                                 \label{inst:0057}
\and Gaia DPAC Project Office, ESAC, Camino bajo del Castillo, s/n, Urbanizacion Villafranca del Castillo, Villanueva de la Ca\~{n}ada, 28692 Madrid, Spain\relax                                            \label{inst:0064}
\and SYRTE, Observatoire de Paris, Universit\'{e} PSL, CNRS,  Sorbonne Universit\'{e}, LNE, 61 avenue de l’Observatoire 75014 Paris, France\relax                                                          \label{inst:0070}
\and National Observatory of Athens, I. Metaxa and Vas. Pavlou, Palaia Penteli, 15236 Athens, Greece\relax                                                                                                   \label{inst:0072}
\and IMCCE, Observatoire de Paris, Universit\'{e} PSL, CNRS,  Sorbonne Universit\'{e}, Univ. Lille, 77 av. Denfert-Rochereau, 75014 Paris, France\relax                                                      \label{inst:0074}
\and Royal Observatory of Belgium, Ringlaan 3, 1180 Brussels, Belgium\relax                                                                                                                                  \label{inst:0075}
\and Institute for Astronomy, University of Edinburgh, Royal Observatory, Blackford Hill, Edinburgh EH9 3HJ, United Kingdom\relax                                                                            \label{inst:0083}
\and Institut d'Astrophysique et de G\'{e}ophysique, Universit\'{e} de Li\`{e}ge, 19c, All\'{e}e du 6 Ao\^{u}t, B-4000 Li\`{e}ge, Belgium\relax                                                              \label{inst:0085}
\and ATG Europe for ESA/ESAC, Camino bajo del Castillo, s/n, Urbanizacion Villafranca del Castillo, Villanueva de la Ca\~{n}ada, 28692 Madrid, Spain\relax                                                   \label{inst:0088}
\and \'{A}rea de Lenguajes y Sistemas Inform\'{a}ticos, Universidad Pablo de Olavide, Ctra. de Utrera, km 1. 41013, Sevilla, Spain\relax                                                                     \label{inst:0092}
\and ETSE Telecomunicaci\'{o}n, Universidade de Vigo, Campus Lagoas-Marcosende, 36310 Vigo, Galicia, Spain\relax                                                                                             \label{inst:0094}
\and Large Synoptic Survey Telescope, 950 N. Cherry Avenue, Tucson, AZ 85719, USA\relax                                                                                                                      \label{inst:0099}
\and Observatoire Astronomique de Strasbourg, Universit\'{e} de Strasbourg, CNRS, UMR 7550, 11 rue de l'Universit\'{e}, 67000 Strasbourg, France\relax                                                       \label{inst:0100}
\and Kavli Institute for Cosmology, University of Cambridge, Madingley Road, Cambride CB3 0HA, United Kingdom\relax                                                                                          \label{inst:0103}
\and Aurora Technology for ESA/ESAC, Camino bajo del Castillo, s/n, Urbanizacion Villafranca del Castillo, Villanueva de la Ca\~{n}ada, 28692 Madrid, Spain\relax                                            \label{inst:0107}
\and Laboratoire Univers et Particules de Montpellier, Universit\'{e} Montpellier, Place Eug\`{e}ne Bataillon, CC72, 34095 Montpellier Cedex 05, France\relax                                                \label{inst:0108}
\and Department of Physics and Astronomy, Division of Astronomy and Space Physics, Uppsala University, Box 516, 75120 Uppsala, Sweden\relax                                                                  \label{inst:0111}
\and CENTRA, Universidade de Lisboa, FCUL, Campo Grande, Edif. C8, 1749-016 Lisboa, Portugal\relax                                                                                                           \label{inst:0112}
\and Universit\`{a} di Catania, Dipartimento di Fisica e Astronomia, Sezione Astrofisica, Via S. Sofia 78, 95123 Catania, Italy\relax                                                                        \label{inst:0113}
\and INAF - Osservatorio Astrofisico di Catania, via S. Sofia 78, 95123 Catania, Italy\relax                                                                                                                 \label{inst:0114}
\and University of Vienna, Department of Astrophysics, T\"{ u}rkenschanzstra{\ss}e 17, A1180 Vienna, Austria\relax                                                                                           \label{inst:0115}
\and CITIC – Department of Computer Science, University of A Coru\~{n}a, Campus de Elvi\~{n}a S/N, 15071, A Coru\~{n}a, Spain\relax                                                                        \label{inst:0117}
\and CITIC – Astronomy and Astrophysics, University of A Coru\~{n}a, Campus de Elvi\~{n}a S/N, 15071, A Coru\~{n}a, Spain\relax                                                                            \label{inst:0118}
\and INAF - Osservatorio Astronomico di Roma, Via di Frascati 33, 00078 Monte Porzio Catone (Roma), Italy\relax                                                                                              \label{inst:0119}
\and Space Science Data Center - ASI, Via del Politecnico SNC, 00133 Roma, Italy\relax                                                                                                                       \label{inst:0120}
\and University of Helsinki, Department of Physics, P.O. Box 64, 00014 Helsinki, Finland\relax                                                                                                               \label{inst:0124}
\and Finnish Geospatial Research Institute FGI, Geodeetinrinne 2, 02430 Masala, Finland\relax                                                                                                                \label{inst:0125}
\and Isdefe for ESA/ESAC, Camino bajo del Castillo, s/n, Urbanizacion Villafranca del Castillo, Villanueva de la Ca\~{n}ada, 28692 Madrid, Spain\relax                                                       \label{inst:0126}
\and Institut UTINAM UMR6213, CNRS, OSU THETA Franche-Comt\'{e} Bourgogne, Universit\'{e} Bourgogne Franche-Comt\'{e}, 25000 Besan\c{c}on, France\relax                                                      \label{inst:0130}
\and STFC, Rutherford Appleton Laboratory, Harwell, Didcot, OX11 0QX, United Kingdom\relax                                                                                                                   \label{inst:0132}
\and Dpto. de Inteligencia Artificial, UNED, c/ Juan del Rosal 16, 28040 Madrid, Spain\relax                                                                                                                 \label{inst:0134}
\and Elecnor Deimos Space for ESA/ESAC, Camino bajo del Castillo, s/n, Urbanizacion Villafranca del Castillo, Villanueva de la Ca\~{n}ada, 28692 Madrid, Spain\relax                                         \label{inst:0142}
\and Thales Services for CNES Centre Spatial de Toulouse, 18 avenue Edouard Belin, 31401 Toulouse Cedex 9, France\relax                                                                                      \label{inst:0143}
\and ON/MCTI-BR, Rua Gal. Jos\'{e} Cristino 77, Rio de Janeiro, CEP 20921-400, RJ,  Brazil\relax                                                                                                             \label{inst:0150}
\and OV/UFRJ-BR, Ladeira Pedro Ant\^{o}nio 43, Rio de Janeiro, CEP 20080-090, RJ, Brazil\relax                                                                                                               \label{inst:0151}
\and Department of Terrestrial Magnetism, Carnegie Institution for Science, 5241 Broad Branch Road, NW, Washington, DC 20015-1305, USA\relax                                                                 \label{inst:0159}
\and Universit\`{a} di Torino, Dipartimento di Fisica, via Pietro Giuria 1, 10125 Torino, Italy\relax                                                                                                        \label{inst:0166}
\and Departamento de Astrof\'{i}sica, Centro de Astrobiolog\'{i}a (CSIC-INTA), ESA-ESAC. Camino Bajo del Castillo s/n. 28692 Villanueva de la Ca\~{n}ada, Madrid, Spain\relax                                \label{inst:0170}
\and Leicester Institute of Space and Earth Observation and Department of Physics and Astronomy, University of Leicester, University Road, Leicester LE1 7RH, United Kingdom\relax                           \label{inst:0172}
\and Departamento de Estad\'{i}stica, Universidad de C\'{a}diz, Calle Rep\'{u}blica \'{A}rabe Saharawi s/n. 11510, Puerto Real, C\'{a}diz, Spain\relax                                                       \label{inst:0177}
\and Astronomical Institute Bern University, Sidlerstrasse 5, 3012 Bern, Switzerland (present address)\relax                                                                                                 \label{inst:0180}
\and EURIX S.r.l., Corso Vittorio Emanuele II 61, 10128, Torino, Italy\relax                                                                                                                                 \label{inst:0181}
\and Harvard-Smithsonian Center for Astrophysics, 60 Garden Street, Cambridge MA 02138, USA\relax                                                                                                            \label{inst:0185}
\and HE Space Operations BV for ESA/ESAC, Camino bajo del Castillo, s/n, Urbanizacion Villafranca del Castillo, Villanueva de la Ca\~{n}ada, 28692 Madrid, Spain\relax                                       \label{inst:0188}
\and Kapteyn Astronomical Institute, University of Groningen, Landleven 12, 9747 AD Groningen, The Netherlands\relax                                                                                         \label{inst:0195}
\and SISSA - Scuola Internazionale Superiore di Studi Avanzati, via Bonomea 265, 34136 Trieste, Italy\relax                                                                                                  \label{inst:0196}
\and University of Turin, Department of Computer Sciences, Corso Svizzera 185, 10149 Torino, Italy\relax                                                                                                     \label{inst:0206}
\and SRON, Netherlands Institute for Space Research, Sorbonnelaan 2, 3584CA, Utrecht, The Netherlands\relax                                                                                                  \label{inst:0207}
\and Dpto. de Matem\'{a}tica Aplicada y Ciencias de la Computaci\'{o}n, Univ. de Cantabria, ETS Ingenieros de Caminos, Canales y Puertos, Avda. de los Castros s/n, 39005 Santander, Spain\relax             \label{inst:0211}
\and Unidad de Astronom\'ia, Universidad de Antofagasta, Avenida Angamos 601, Antofagasta 1270300, Chile\relax                                                                                               \label{inst:0218}
\and CRAAG - Centre de Recherche en Astronomie, Astrophysique et G\'{e}ophysique, Route de l'Observatoire Bp 63 Bouzareah 16340 Algiers, Algeria\relax                                                       \label{inst:0229}
\and University of Antwerp, Onderzoeksgroep Toegepaste Wiskunde, Middelheimlaan 1, 2020 Antwerp, Belgium\relax                                                                                               \label{inst:0232}
\and INAF - Osservatorio Astronomico d'Abruzzo, Via Mentore Maggini, 64100 Teramo, Italy\relax                                                                                                               \label{inst:0234}
\and INAF - Osservatorio Astronomico di Capodimonte, Via Moiariello 16, 80131, Napoli, Italy\relax                                                                                                           \label{inst:0236}
\and Instituto de Astronomia, Geof\`{i}sica e Ci\^{e}ncias Atmosf\'{e}ricas, Universidade de S\~{a}o Paulo, Rua do Mat\~{a}o, 1226, Cidade Universitaria, 05508-900 S\~{a}o Paulo, SP, Brazil\relax          \label{inst:0237}
\and Department of Astrophysics, Astronomy and Mechanics, National and Kapodistrian University of Athens, Panepistimiopolis, Zografos, 15783 Athens, Greece\relax                                            \label{inst:0248}
\and Leibniz Institute for Astrophysics Potsdam (AIP), An der Sternwarte 16, 14482 Potsdam, Germany\relax                                                                                                    \label{inst:0251}
\and RHEA for ESA/ESAC, Camino bajo del Castillo, s/n, Urbanizacion Villafranca del Castillo, Villanueva de la Ca\~{n}ada, 28692 Madrid, Spain\relax                                                         \label{inst:0253}
\and ATOS for CNES Centre Spatial de Toulouse, 18 avenue Edouard Belin, 31401 Toulouse Cedex 9, France\relax                                                                                                 \label{inst:0255}
\and UNINOVA - CTS, Campus FCT-UNL, Monte da Caparica, 2829-516 Caparica, Portugal\relax                                                                                                                     \label{inst:0259}
\and School of Physics, O'Brien Centre for Science North, University College Dublin, Belfield, Dublin 4, Ireland\relax                                                                                       \label{inst:0270}
\and Dipartimento di Fisica e Astronomia, Universit\`{a} di Bologna, Via Piero Gobetti 93/2, 40129 Bologna, Italy\relax                                                                                      \label{inst:0276}
\and Barcelona Supercomputing Center - Centro Nacional de Supercomputaci\'{o}n, c/ Jordi Girona 29, Ed. Nexus II, 08034 Barcelona, Spain\relax                                                               \label{inst:0287}
\and Department of Computer Science, Electrical and Space Engineering, Lule\aa{} University of Technology, Box 848, S-981 28 Kiruna, Sweden\relax                                                            \label{inst:0292}
\and Max Planck Institute for Extraterrestrial Physics, High Energy Group, Gie{\ss}enbachstra{\ss}e, 85741 Garching, Germany\relax                                                                           \label{inst:0294}
\and Astronomical Observatory Institute, Faculty of Physics, Adam Mickiewicz University, S{\l}oneczna 36, 60-286 Pozna{\'n}, Poland\relax                                                                    \label{inst:0313}
\and E\"{ o}tv\"{ o}s Lor\'and University, Egyetem t\'{e}r 1-3, H-1053 Budapest, Hungary\relax                                                                                                               \label{inst:0320}
\and American Community Schools of Athens, 129 Aghias Paraskevis Ave. \& Kazantzaki Street, Halandri, 15234 Athens, Greece\relax                                                                             \label{inst:0323}
\and Faculty of Mathematics and Physics, University of Ljubljana, Jadranska ulica 19, 1000 Ljubljana, Slovenia\relax                                                                                         \label{inst:0326}
\and Villanova University, Department of Astrophysics and Planetary Science, 800 E Lancaster Avenue, Villanova PA 19085, USA\relax                                                                           \label{inst:0327}
\and Physics Department, University of Antwerp, Groenenborgerlaan 171, 2020 Antwerp, Belgium\relax                                                                                                           \label{inst:0329}
\and McWilliams Center for Cosmology, Department of Physics, Carnegie Mellon University, 5000 Forbes Avenue, Pittsburgh, PA 15213, USA\relax                                                                 \label{inst:0335}
\and Astronomical Institute, Academy of Sciences of the Czech Republic, Fri\v{c}ova 298, 25165 Ond\v{r}ejov, Czech Republic\relax                                                                            \label{inst:0339}
\and Telespazio for CNES Centre Spatial de Toulouse, 18 avenue Edouard Belin, 31401 Toulouse Cedex 9, France\relax                                                                                           \label{inst:0344}
\and Institut de Physique de Rennes, Universit{\'e} de Rennes 1, 35042 Rennes, France\relax                                                                                                                  \label{inst:0347}
\and Shanghai Astronomical Observatory, Chinese Academy of Sciences, 80 Nandan Rd, 200030 Shanghai, China\relax                                                                                              \label{inst:0354}
\and School of Astronomy and Space Science, University of Chinese Academy of Sciences, Beijing 100049, China\relax                                                                                           \label{inst:0355}
\and Niels Bohr Institute, University of Copenhagen, Juliane Maries Vej 30, 2100 Copenhagen {\O}, Denmark\relax                                                                                              \label{inst:0357}
\and DXC Technology, Retortvej 8, 2500 Valby, Denmark\relax                                                                                                                                                  \label{inst:0358}
\and Las Cumbres Observatory, 6740 Cortona Drive Suite 102, Goleta, CA 93117, USA\relax                                                                                                                      \label{inst:0359}
\and Astrophysics Research Institute, Liverpool John Moores University, 146 Brownlow Hill, Liverpool L3 5RF, United Kingdom\relax                                                                            \label{inst:0367}
\and Baja Observatory of University of Szeged, Szegedi \'{u}t III/70, 6500 Baja, Hungary\relax                                                                                                               \label{inst:0372}
\and Laboratoire AIM, IRFU/Service d'Astrophysique - CEA/DSM - CNRS - Universit\'{e} Paris Diderot, B\^{a}t 709, CEA-Saclay, 91191 Gif-sur-Yvette Cedex, France\relax                                        \label{inst:0373}
\and \'{E}cole polytechnique f\'{e}d\'{e}rale de Lausanne, SB MATHAA STAP, MA B1 473 (B\^{a}timent MA), Station 8, CH-1015 Lausanne, Switzerland\relax                                                       \label{inst:0390}
\and Warsaw University Observatory, Al. Ujazdowskie 4, 00-478 Warszawa, Poland\relax                                                                                                                         \label{inst:0411}
\and Institute of Theoretical Physics, Faculty of Mathematics and Physics, Charles University in Prague, Czech Republic\relax                                                                                \label{inst:0412}
\and AKKA for CNES Centre Spatial de Toulouse, 18 avenue Edouard Belin, 31401 Toulouse Cedex 9, France\relax                                                                                                 \label{inst:0419}
\and Vitrociset Belgium for ESA/ESAC, Camino bajo del Castillo, s/n, Urbanizacion Villafranca del Castillo, Villanueva de la Ca\~{n}ada, 28692 Madrid, Spain\relax                                           \label{inst:0425}
\and HE Space Operations BV for ESA/ESTEC, Keplerlaan 1, 2201AZ, Noordwijk, The Netherlands\relax                                                                                                            \label{inst:0433}
\and Space Telescope Science Institute, 3700 San Martin Drive, Baltimore, MD 21218, USA\relax                                                                                                                \label{inst:0442}
\and QUASAR Science Resources for ESA/ESAC, Camino bajo del Castillo, s/n, Urbanizacion Villafranca del Castillo, Villanueva de la Ca\~{n}ada, 28692 Madrid, Spain\relax                                     \label{inst:0443}
\and Fork Research, Rua do Cruzado Osberno, Lt. 1, 9 esq., Lisboa, Portugal\relax                                                                                                                            \label{inst:0448}
\and APAVE SUDEUROPE SAS for CNES Centre Spatial de Toulouse, 18 avenue Edouard Belin, 31401 Toulouse Cedex 9, France\relax                                                                                  \label{inst:0451}
\and Nordic Optical Telescope, Rambla Jos\'{e} Ana Fern\'{a}ndez P\'{e}rez 7, 38711 Bre\~{n}a Baja, Spain\relax                                                                                              \label{inst:0456}
\and Spanish Virtual Observatory\relax                                                                                                                                                                       \label{inst:0461}
\and Fundaci\'{o}n Galileo Galilei - INAF, Rambla Jos\'{e} Ana Fern\'{a}ndez P\'{e}rez 7, 38712 Bre\~{n}a Baja, Santa Cruz de Tenerife, Spain\relax                                                          \label{inst:0473}
\and INSA for ESA/ESAC, Camino bajo del Castillo, s/n, Urbanizacion Villafranca del Castillo, Villanueva de la Ca\~{n}ada, 28692 Madrid, Spain\relax                                                         \label{inst:0477}
\and Dpto. Arquitectura de Computadores y Autom\'{a}tica, Facultad de Inform\'{a}tica, Universidad Complutense de Madrid, C/ Prof. Jos\'{e} Garc\'{i}a Santesmases s/n, 28040 Madrid, Spain\relax            \label{inst:0478}
\and H H Wills Physics Laboratory, University of Bristol, Tyndall Avenue, Bristol BS8 1TL, United Kingdom\relax                                                                                              \label{inst:0481}
\and Institut d'Estudis Espacials de Catalunya (IEEC), Gran Capita 2-4, 08034 Barcelona, Spain\relax                                                                                                         \label{inst:0487}
\and Applied Physics Department, Universidade de Vigo, 36310 Vigo, Spain\relax                                                                                                                               \label{inst:0489}
\and Stellar Astrophysics Centre, Aarhus University, Department of Physics and Astronomy, 120 Ny Munkegade, Building 1520, DK-8000 Aarhus C, Denmark\relax                                                   \label{inst:0504}
\and Argelander-Institut f\"{ ur} Astronomie, Universit\"{ a}t Bonn,  Auf dem H\"{ u}gel 71, 53121 Bonn, Germany\relax                                                                                       \label{inst:0510}
\and Research School of Astronomy and Astrophysics, Australian National University, Canberra, ACT 2611 Australia\relax                                                                                       \label{inst:0517}
\and Sorbonne Universit\'{e}s, UPMC Univ. Paris 6 et CNRS, UMR 7095, Institut d'Astrophysique de Paris, 98 bis bd. Arago, 75014 Paris, France\relax                                                          \label{inst:0519}
\and Department of Geosciences, Tel Aviv University, Tel Aviv 6997801, Israel\relax                                                                                                                          \label{inst:0521}
}

\date{Received ; accepted } 

\abstract
   {
   The ESA \gaia mission provides a unique time-domain survey for more than 
   $1.6$~billion sources with $G \lesssim 21$~mag.
   }
%
   {
   We showcase stellar variability in the Galactic colour-absolute magnitude diagram (CaMD). We focus on pulsating, eruptive, and cataclysmic variables, as well as on stars that exhibit variability that is due to rotation and eclipses. 
   }
   {
   We describe the locations of variable star classes, variable object fractions, and typical variability amplitudes throughout the CaMD and show how variability-related changes in colour and brightness induce `motions'. To do this, we
   use 22 months of calibrated photometric, spectro-photometric, and astrometric \gaia data of stars with a significant parallax.

   To ensure that a large variety of variable star classes populate the CaMD, we crossmatched \gaia sources 
   with known variable stars. 
   We also used the statistics and variability detection modules of the \gaia variability pipeline.
   Corrections for interstellar extinction are not implemented in this article.
   }
%
   {
   \gaia enables the first investigation of Galactic variable star populations in the CaMD on a similar, if not larger, scale as was previously done in the Magellanic Clouds. Although the observed colours are not corrected for reddening,
   distinct regions are visible in which variable stars occur. We determine variable star fractions 
   to within the current detection thresholds of {\it Gaia}.
   Finally, we report the most complete description of variability-induced motion within the CaMD to date. 
   }
%
   {
   \gaia enables novel insights into variability phenomena 
   for an unprecedented number of stars, which will benefit the understanding of stellar astrophysics. The CaMD of Galactic variable stars provides crucial information on physical origins of variability in a way that has previously only been accessible for Galactic star clusters or external galaxies.
   Future \gaia data releases will enable significant improvements over this preview by providing longer time series, more accurate astrometry, and additional data types (time series BP and RP spectra, RVS spectra, and radial velocities),
   all for much larger samples of stars. 
   }


\maketitle

\section{Introduction \label{sect:introduction}}
The ESA space mission \gaia \citep{DPACP-1} has been conducting a unique survey since the beginning of its operations (end of July 2014). Its uniqueness derives from several aspects that we list in the following paragraphs.

Firstly, \gaia delivers nearly simultaneous measurements in the three observational domains on which most stellar astronomical studies are based: astrometry, photometry, and spectroscopy \citep{Brown16,DPACP-12}. As consequence of the spin of the spacecraft, it takes about 80 seconds for sources to be measured from the first to the last CCD during a single field-of-view transit.

Secondly, the \gaia data releases provide accurate astrometric measurements  
for an unprecedented number of objects. In particular, trigonometric parallaxes carry invaluable information, since they are required to infer stellar luminosities, which form the basis of the understanding of much of stellar astrophysics. 
Proper and orbital motions of stars further enable mass measurements in multiple stellar systems, as well as the investigation of cluster membership.

Thirdly, \gaia data are homogeneous throughout the entire sky, since they are observed with a single set of instruments and are not subject to the Earth's
atmosphere
or seasons. 
All-sky surveys cannot be achieved using a single ground-based telescope; 
surveys using multiple sites and telescopes and instruments require cross-calibration, which unavoidably introduces systematics and reduces precision because of the increased scatter. Thus, \gaia will play an important role as a standard source in cross-calibrating heterogeneous surveys and instruments, much like the \textit{Hipparcos} mission \citep{Perryman1997Hipparcos,HIPPARCOS_PERIODIC_ESA_1997} did in the past. Of course, \gaia represents a quantum leap from {\it Hipparcos} in many regards, including an increase of four orders of magnitude in the number of objects observed, additional types of observations (spectrophotometry and spectroscopy), and significantly improved sensitivity and precision for all types of measurements.

Fourthly, there are unprecedented synergies for calibrating distance scales using the{\it } dual astrometric and time-domain capabilities of \textit{Gaia}  \citep[e.g.][]{EyerEtAl2012}. Specifically, \gaia will enable the discovery of unrivalled numbers of standard candles residing in the Milky Way, and anchor Leavitt laws (period-luminosity relations) to trigonometric parallaxes \citep[see][for two examples based on the first \gaia data release]{GaiaDR1_CepRRL,2017A&A...599A..67C}.

Variable stars have for a long time been recognised to offer crucial insights 
into stellar structure and evolution. Similarly, the Hertzsprung-Russell diagram (HRD) provides an overview of all stages of stellar evolution, and together with its empirical cousin, the colour-magnitude diagram (CMD), it has shaped stellar astrophysics like no other diagram. Henrietta Leavitt (1908) was one of the first to note the immense potential of studying variable stars in populations, where distance 
uncertainties
did not introduce significant scatter. Soon thereafter, \citet{1912HARCI.173....1l} discovered the period-luminosity relation of Cepheid variables, which has become a cornerstone of stellar physics and cosmology. It appears that \citet[his Fig.\,42]{1951ApJ...113..367E} was the first to use (photoelectric) observations of variable stars (in this case, classical Cepheids) to constrain regions where Cepheids occur in the HRD; these regions are today referred to as instability strips. Eggen further illustrated how Cepheids change their locus in the colour-absolute magnitude diagram (CaMD) during the course of their variability, thus developing a time-dependent CMD for variable stars. \citet{1956PZ.....11..325K} and \citet{1958ApJ...128..150S} later illustrated the varying locations of variable stars in the HRD using classical Cepheids located within star clusters. By combining the different types of \gaia time-series data with \gaia parallaxes, we are now in a position to construct time-dependent CaMD towards any direction in the Milky Way, building on previous work based on {\it Hipparcos} \citep{EyerEtAl1994,EyerGrenon1997}, but on a much larger scale.

Many variability (ground- and space-based) surveys have exploited the power of identifying variable stars in stellar populations at similar distances, for example, in star clusters or nearby galaxies such as the Magellanic Clouds. Ground-based microlensing surveys such as the Optical Gravitational Lensing Experiment \citep[OGLE; e.g.][]{2015AcA....65....1U}, the Exp\'erience pour la Recherche d'Objets Sombres \citep{EROS1999}, and the Massive Compact Halo Objects project \citep[MACHO;][]{Alcock1993MACHO} deserve a special mention in this regard. 
The data will continue to grow with the next large multi-epoch surveys such as the Zwicky Transient Facility \citep{2014htu..conf...27B} and the Large Synoptic Survey Telescope \citep{2009arXiv0912.0201L} from the ground, and the Transiting Exoplanet Survey Satellite \citep[TESS;][]{2015JATIS...1a4003R} and PLATO \citep{2014ExA....38..249R} from space. 

Another ground-breaking observational trend has been the long-term high-precision high-cadence uninterrupted space photometry with CoRoT/BRITE \citep[][with time bases of up to five months]{2009A&A...506..411A,Pablo2016} and {\it Kepler\/}/K2 \citep[][with time bases of up to four years and three months, respectively]{Gilliland2010,Howell2014} provided entirely new insights into micro-magnitude level variability of stars, with periodicities ranging from minutes to years. 
These missions opened up stellar interiors from the detection of solar-like oscillations of thousands of sun-like stars and red giants
\citep[e.g.][for reviews]{Bedding2011,ChaplinMiglio2013,HekkerJCD2017}, as well as hundreds of intermediate-mass stars \citep[e.g.][]{Aerts2015,Bowman2017} and compact pulsators \citep[e.g.][]{Hermes2017}.
The results we provide in Sects.\,3 and 4 on the variability fractions and levels are representative of milli-mag level variability and not of micro-mag levels, as are found in space asteroseismic data.

Any of these asteroseismic surveys can benefit from \gaia astrometry, however, so that distances and luminosities can be derived, as \cite{DeRidder2016} and \cite{2017ApJ...844..102H} reported with \gaia DR1 data. \gaia will also contribute to these surveys with its photometry, and some surveys will also benefit from the \gaia radial velocities (depending on their operating magnitude range).

\begin{figure*}[h!]
\centering
 \includegraphics[width=180mm]{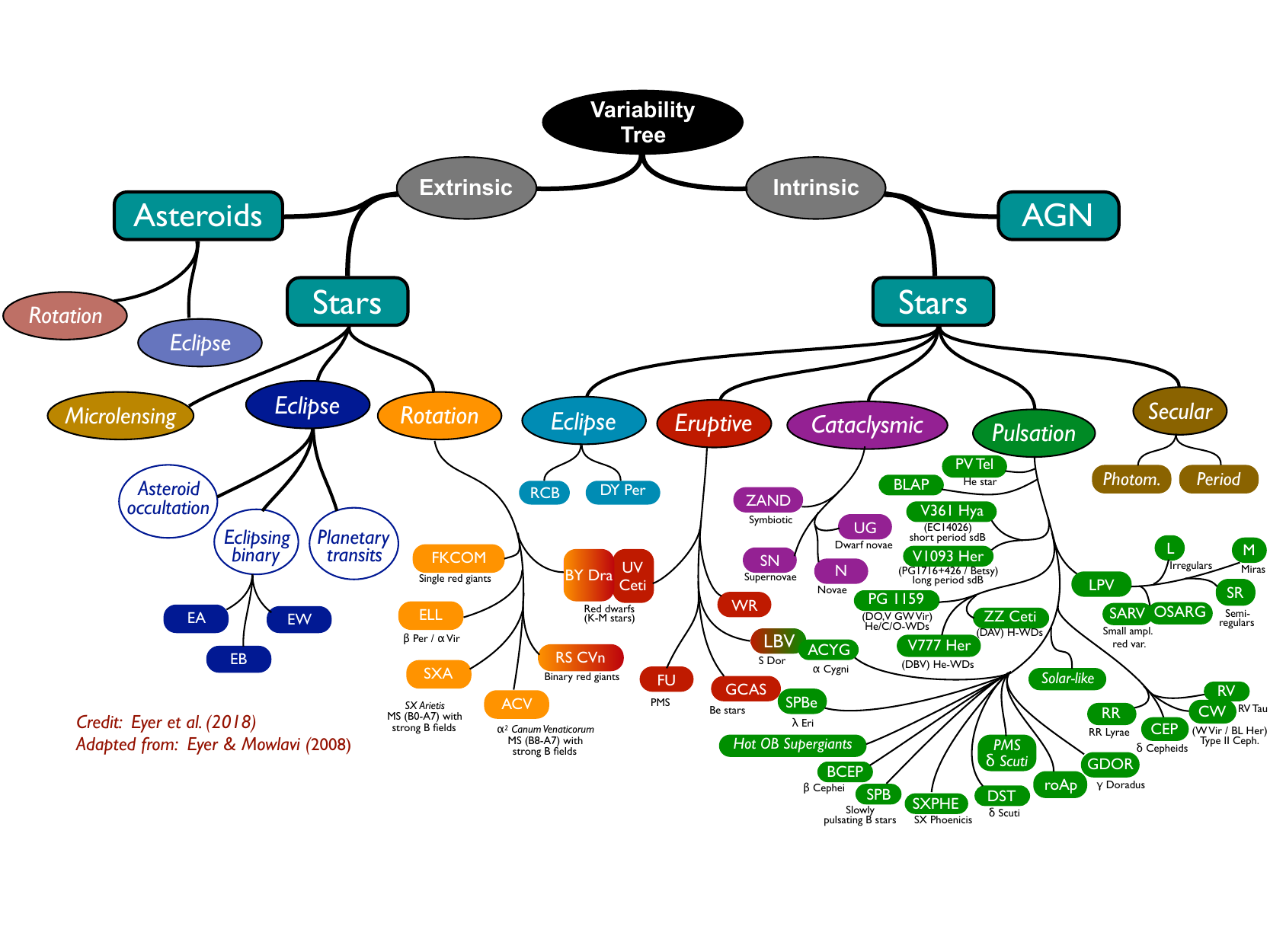}
 \caption{Updated version of the variability tree presented in \cite{EyerMowlavi2008}, separated according to the cause of
 variability phenomena: variability induced by pulsations, rotation, eruptions, eclipses, and cataclysmic events.
 }
 \label{fig:VariabilityTree}
\end{figure*}

Stellar variability 
comprises a great variety of observable features that are due to different physical origins.
Figure\,\ref{fig:VariabilityTree} shows the updated variability tree \citep{EyerMowlavi2008}, which provides a useful overview of the various types of variability and their known causes.  
The variability tree has four levels: the distinction of intrinsic versus extrinsic variability, 
the separation into major types of objects (asteroid, stars, and AGN), the physical origin of the variability, and the class name. 
In this article, we follow the classical distinction of the different causes of the variability phenomena: 
variability induced by pulsation, rotation, eruption, eclipses, and cataclysmic events.
A large number of variability types can be identified in the \gaia data even now, as described in the subsequent sections.

We here provide an overview of stellar variability in the 
CaMD, building on the astrometric and photometric data of the second \gaia data release (DR2).
Future \gaia DRs will enable much more detailed investigations of this kind using longer temporal baselines, greater number of observations, and added classes of variable stars (such as eclipsing binaries, which will be published in DR3).

This paper is structured as follows. 
Section~\ref{sect:location} shows the location of different variability types in the CaMD, making use of known objects from the literature that are published in \gaia DR2, but without any further analysis of the \gaia data. 
Section~\ref{sect:fraction} presents the fraction of variables as a function of colour and absolute magnitude, obtained by processing the \gaia time series for the detection of variability \citep{EyerGaiaDoc2018}.
Section~\ref{sect:variationlevel} investigates the variability level in the CaMD by employing statistics and classification results (some of which are related to unpublished \gaia time series).
Section~\ref{sect:motion} shows the motion of known variables stars in the CaMD, that is, a \textup{time-dependent CaMD}, which also includes sources that are not available in the DR2 archive but are online material.
Section~\ref{sect:conclusions} summarises our results and presents an outlook to future \gaia DRs. Further information on the literature cross-match and on the selection criteria applied to our data samples can be found in Appendices \ref{app:literature} and \ref{app:selection}, respectively.

\section{Location of variability types in the CaMD \label{sect:location} }
The precision of the location in the CaMD depends on the precision of the colour on the one hand and on the determination of the absolute magnitude on the other. The precision of the absolute magnitude of variable stars
depends on  the photometric precision, the number of measurements, the amplitude of variability, and the relative parallax precision 
$\sigma_{\varpi}/\varpi$. The upper limits of $\sigma_{\varpi}/\varpi$ employed in this article vary between 5 and 20\%, which means that the uncertainty of the absolute magnitude that is solely due to the parallax uncertainty can be as large as $5\,(\ln 10)^{-1} \sigma_{\varpi}/\varpi \approx 0.43$~mag.

As we determined the colour as a function of integrated BP and integrated RP spectro-photometric measurements with tight constraints on the precision of these quantities (see Appendix~\ref{app:selection}), there are parts of the CaMD that are not explored here. For example, the faint end of the main sequence presented in Fig.~9 of \cite{DPACP-31} 
does not fulfil the condition on the precision in BP, so our diagrams do not include L and T brown dwarfs (which are fainter than $M_G \sim 14$~mag). If we cross-match the \gaia data (conditional on the selection of Appendix~\ref{app:selection}) with the catalogue of M dwarfs of \cite{LepinGaidos2011}, only a few M6, M7, and M8 dwarf stars are found.

\begin{figure*}[th!]
\centering
 \includegraphics[width=180mm]{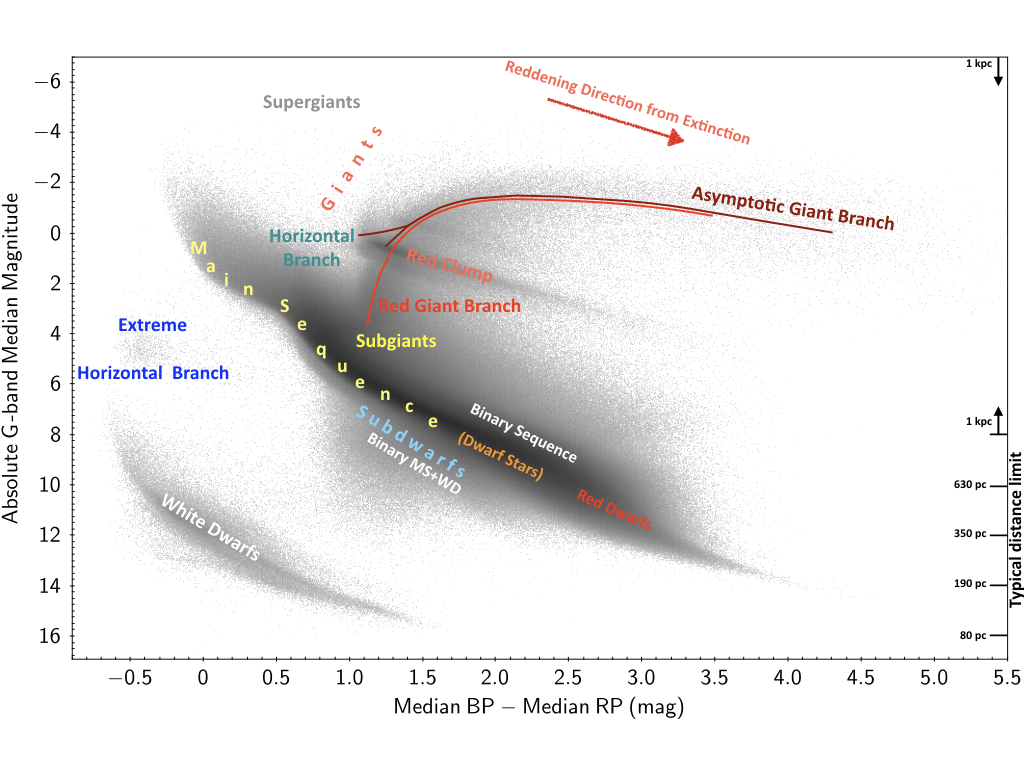}
   \caption{CaMD with its most striking known features (see text). The points in grey denote objects with parallax greater than 1\,mas, with relative parallax precision better than 20\% and other criteria described in Appendix~\ref{app:selection}.} 
    \label{fig:CommentedCMD}
\end{figure*}

In Fig.~\ref{fig:CommentedCMD} we introduce the Gaia CaMD, which is displayed as a background in subsequent figures. For clarity, we note basic astronomical features such as the main sequence, the red clump (and its long tail due to interstellar extinction), the horizontal branch, the extreme horizontal branch \citep[see][for its physical origin]{DCruz1996}, the red giant branch, the asymptotic giant branch, the white dwarf sequence, the subdwarfs, the supergiants, and the binary sequence.
There are additional subtle features above and below the red clump that are described in Fig.~10 of \citet{DPACP-31} and are known as the asymptotic giant branch bump and the red giant branch bump, respectively.
On the right-hand side of Fig.~\ref{fig:CommentedCMD}, we also note the typical limiting distance that can be reached because of the selection of $\sigma_{\varpi}/\varpi$, up to 1 kpc (which was the largest distance we considered for background stars). 

Several effects can influence the average location of a star in the CaMD (in both axes), including interstellar extinction, stellar multiplicity, rotation, inclination of the rotation axis, and chemical composition. In this work, we do not correct for such phenomena and instead rely on the apparent magnitudes and colours measured by \gaia, computing `absolute' magnitudes using \gaia parallaxes. We note that interstellar extinction and reddening can be significant at the considered distances (up to $1$\,kpc), in particular for objects in the Galactic plane. This leads to distortions of certain observed features, such as the long tail in the red clump, which extends to redder and fainter magnitudes.

The stellar variability aspects covered in the second Data Release of \gaia include a limited 
number of variability classes \citep{DPACP-49}: long-period variables, Cepheids, RR\,Lyrae stars, SX\,Phoenicis/$\delta$\,Scuti stars, and rotation-modulated solar-like variability (i.e. all late-type BY\,Draconis stars).
Short-timescale variability (within one day)
was explored
regardless of the physical origin of the variability \citep{Roelens_2018}, although stars that are classified as eclipsing binaries were removed as planned to first appear in the third Data Release of \gaia. The stars presented in this section are solely based on the  cross-match with known objects in the literature. The list of variability types presented here is not meant to be comprehensive.

Figures~\ref{fig:varTypeCMD_Puls}--\ref{fig:varTypeCMD_Catac} illustrate the locations of known variable stars from catalogues in the literature that are cross-matched with the \gaia data.
We indicate these targets according to their known variability type published in the literature (the references are listed in Table~\ref{tab:literature}), and only the stars that satisfy the selection criteria described in Appendix~\ref{app:selection} are kept.
Each of these figures includes as reference the location and density (in grey scale) of all stars, regardless of stellar variability, that satisfy the astrometric and photometric criteria of Appendix~\ref{app:selection} with the additional constraint of a minimum parallax of 1~mas (i.e. within 1 kpc to the Sun). This radius seems a good compromise between a large number of stars and a limited effect of interstellar matter. 
Variable stars whose variability type was previously known 
are represented by combinations of symbols and colours.
Following the structure of the variability tree in Fig.~\ref{fig:VariabilityTree}, we show in separate figures the CaMDs of stars whose variability is induced by different causes, such as pulsations, rotation, eruptions, eclipses, and cataclysmic events.

Several caveats apply to Figs.~\ref{fig:varTypeCMD_Puls}--\ref{fig:varTypeCMD_Catac} and should be kept in mind for their interpretation. 
(a)~The quality of catalogues published in the literature can be rather 
different,
in part because variability is often classified without knowledge of a parallax. To reduce the impact of misclassified objects on these figures, we selected subsets of all available catalogues as reference for specific variable star classes, depending on their agreement with the expected locations in the CaMD.
In certain cases, we have 
excluded sources from the literature by choice of specific catalogues (Table~\ref{tab:literature}) and by using the \gaia astrometry and the multi-band photometric time-series data for occasional cuts in magnitude or colour. 
Future \gaia data releases will provide a more homogeneous variability classification that will rely primarily on the results of the variability processing \citep{DPACP-49}.
(b)~The CaMDs are not corrected for  extinction, which leads to increased scatter, in particular for objects that primarily reside in heavily attenuated areas such as the Galactic disc and the Galactic bulge.
(c)~The cross-match of sources can be erroneous when stars are located in crowded regions or have high proper motion,
especially if the positions of stars in the published catalogues are not sufficiently precise or if proper motion information is not available. 
(d)~Some variability types like magnetically active stars (e.g. RS~CVn stars) exhibit different observational phenomena, such as rotational modulation variations as well as flares. To avoid overcrowding the CaMD diagrams, these types are represented in only one of the relevant diagrams (e.g. with rotational or eruptive variables). Furthermore, we note that the time sampling and the waveband coverage of a given survey might favour the detection of only some of these aspects.
(e)~\gaia represents a milestone for space astrometry and photometry. Nevertheless, some sources can be affected by problems such as corrupt measurements
so that their location in the CaMD may be incorrect \citep{DPACP-39}. 
However, we stress that these problems are limited to a small fraction of sources so that most known variable classes are recovered as expected. The cyclic approach of the \gaia data processing and analysis will allow us to correct for these unexpected features in the future data releases.

\subsection{Pulsating variable stars\label{sec:pulsationCaMD}}

\begin{figure*}[th!]
\centering
 \includegraphics[width=180mm]{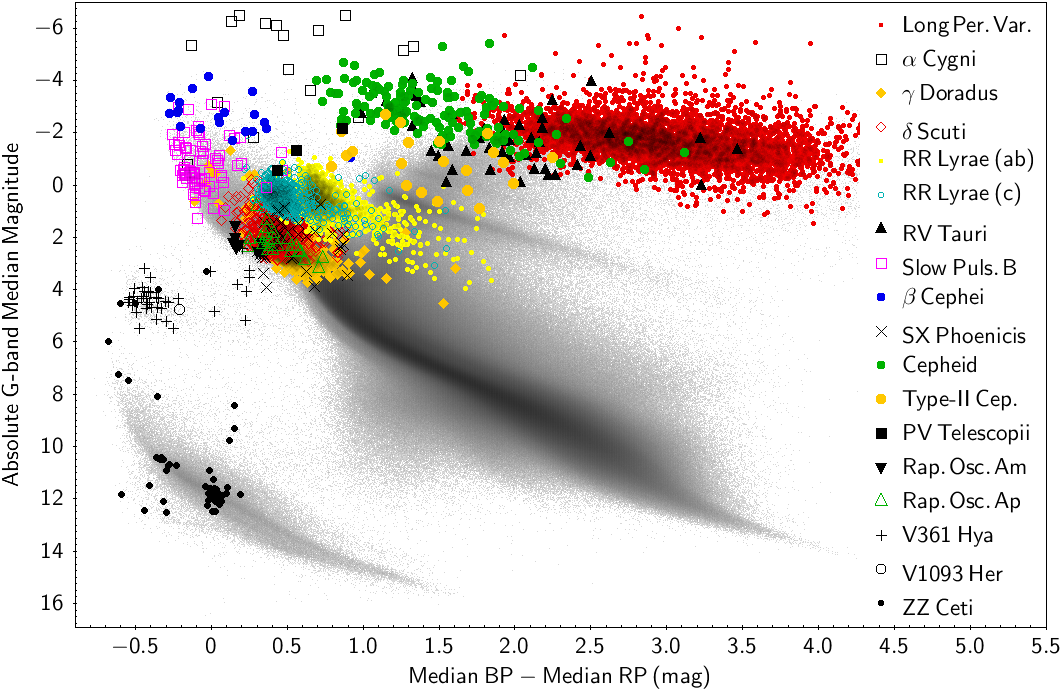}
   \caption{Known pulsating variable stars retrieved from published catalogues are placed in the observational CaMD, with symbols and colours representing types as shown in the legend (see Appendix~\ref{tab:literature} for the references from the literature per type). All stars satisfy the selection criteria described in Appendix~\ref{app:selection}. The background points in
      grey denote a reference subset of objects with a stricter constraint on parallax ($\varpi > 1$\,mas), which limits the sample size, extinction, and reddening. The effects of interstellar matter and other phenomena (see text) are not corrected for.
      The condition on the relative precision of \gbp measurements introduces artificial cuts in the distributions of low-mass main-sequence stars and red (super)giants.} 
    \label{fig:varTypeCMD_Puls}
\end{figure*}

Figure~\ref{fig:varTypeCMD_Puls} shows the positions of different classes of pulsating variable stars based on the \gaia data and can be compared to its theoretical counterpart  
in recent textbooks on asteroseismology \citep[Fig.~1.12 in][]{2010aste.book.....A} and on pulsating stars \citep{2015pust.book.....C}. We refer to these books for further details of specific variability classes. 
Here, we only consider the following types of pulsating variable stars: 
\begin{enumerate}
        \item Long-period variables, which are red giant stars that populate the reddest and brightest regions of the CaMD. They include Miras, semi-regular variables, slow irregular variables, and small-amplitude red giants.  
        
        \item $\alpha$ Cygni stars, which are luminous supergiant stars that pulsate in non-radial modes. They are particularly affected by interstellar extinction as they are young massive stars that reside in the Galactic disc, so that their position in Fig.~\ref{fig:varTypeCMD_Puls} must be treated with caution.
        
        \item $\delta$\,Scuti stars, which are Population-I stars of spectral types A and F with short periods ($<0.3$ d) that dominantly pulsate in pressure modes, but may also reveal low-order gravity modes of low amplitude.
        
        \item SX Phoenicis stars, which are Population-II high-amplitude $\delta$\,Scuti stars. 

        \item $\gamma$\,Doradus stars, which are stars with spectral types A and F with periods from 0.3 to 3\,d that  dominantly pulsate in high-order gravity modes, but may also reveal low-amplitude pressure modes.
        \item RR Lyrae stars (Bailey's type ab and c), which are Population-II horizontal branch stars with periods from 0.2 to 1\,d that pulsate in pressure mode. 
        C-type RR~Lyrae stars are bluer than ab-type stars.
        \item Slowly pulsating B (SPB) stars, which are non-radial multi-periodic gravity-mode pulsators of spectral type B with periods typically in the range from 0.4 to 5\,d.
        \item $\beta$ Cephei stars, which are late-O to early-B spectral type non-supergiant stars with dominant low-order pressure and gravity modes that feature periods in the range from 0.1 to 0.6\,d. Several of them have been found to also exhibit low-amplitude high-order gravity modes as in the SPB stars \citep[e.g.][]{2005ApJS..158..193S}. 
        The $\beta$~Cephei stars are located in the Galactic disc so that their CaMD position is easily affected by interstellar extinction.

        \item Classical Cepheids (prototype $\delta$\,Cephei), which are evolved Population-I (young intermediate-mass) stars featuring radial p-mode pulsations with periods of approximately $1-100$\,d. Cepheids can be strongly affected by interstellar extinction as they reside in the Galactic disc and can be observed at great distances.
        
        \item Type-II Cepheids, which are Population-II stars pulsating in p-mode that were historically thought to be identical to classical Cepheids. Type-II Cepheids consist of three different sub-classes (separated by period) that are commonly referred to as BL~Herculis, W~Virginis, and RV~Tauri stars. Their evolutionary scenarios differ significantly, although the three sub-classes together define a tight period-luminosity relation.
        \item PV Telescopii stars, which include the sub-classes V652~Her, V2076~Oph, and FQ~Aqr \citep{2008IBVS.5817....1J}. They are rare  hydrogen-deficient supergiant stars that cover a wide range of spectral types  and exhibit complex light and radial velocity variations.
        \item Rapidly oscillating Am and Ap stars, which are chemically peculiar A stars that exhibit multiperiodic non-radial pressure modes in the period range of about $5-20$\,min.
        \item V361\,Hydrae (or EC\,14026) stars, which are subdwarf B stars on the extreme horizontal branch that pulsate in pressure modes with very short periods of $\sim 1 - 10$\,min.
        \item V1093\,Her (or PG\,1716) stars, which are subdwarf B stars on the extreme horizontal branch that pulsate in gravity modes with periods of about $1 - 3$\,h.
        \item ZZ Ceti stars, which are white dwarfs featuring fast non-radial gravity-mode pulsations with periods of $0.5 - 25$\,min. 
\end{enumerate}

The CaMD of pulsating stars carries a great deal of information, much of which has shaped the understanding of stellar structure and evolution and can be found in textbooks. Briefly summarised, we note the following particularly interesting features of Fig.\,\ref{fig:varTypeCMD_Puls}.
\begin{itemize}
    \item Extinction affects variability classes that belong to different populations unequally, as expected. Stars located away from the Galactic disc are much less reddened and thus clump more clearly. This effect is particularly obvious when RR~Lyrae stars and classical Cepheids are compared, which both occupy the same instability strip, and it cannot be explained by the known fact that the classical instability strip becomes wider in colour at higher luminosity \citep[e.g. see][and references therein]{2016A&A...591A...8A, 2005ApJ...632..590M, 2000ApJ...529..293B}. 
    \item Interstellar reddening blurs the boundaries between variability classes. Correcting for interstellar extinction will be crucial to delineate the borders of the instability strips in the CaMD, as well as to deduce their purity in terms of the fraction of stars that exhibit pulsations while residing in such regions.
    \item Practical difficulties involved in separating variable star classes in the way required to construct  Fig.\,\ref{fig:varTypeCMD_Puls} include a) that variable stars are often subject to multiple types of variability (e.g. $\gamma$~Doradus/$\delta$~Scuti, $\beta$~Cephei/SPB hybrid pulsators, pulsating stars in eclipsing binary systems, or pulsating white dwarfs that exhibit eruptions), and b) that naming conventions are often historical or purely based on light-curve morphology, so that they do not account for different evolutionary scenarios (e.g. type-II Cepheids). With additional data and a fully homogeneous variable star classification based on \gaia alone, such ambiguities will be resolved in the future unless they are intrinsically connected to the nature of the variability.
    \item We note multiple groups of ZZ~Ceti stars along the white dwarf sequence. The most prominent of these is located at $\mbox{\gbp}-\mbox{\grp} \simeq 0$ and $M_G \simeq 12,$ as reported in \cite{FontaineBrassard2008}.
\end{itemize}

\subsection{Variability due to rotation and eclipses\label{sec:rotationCaMD}\label{sec:eclipsesCaMD}}

\begin{figure*}
\centering
 \includegraphics[height=0.45\textheight]{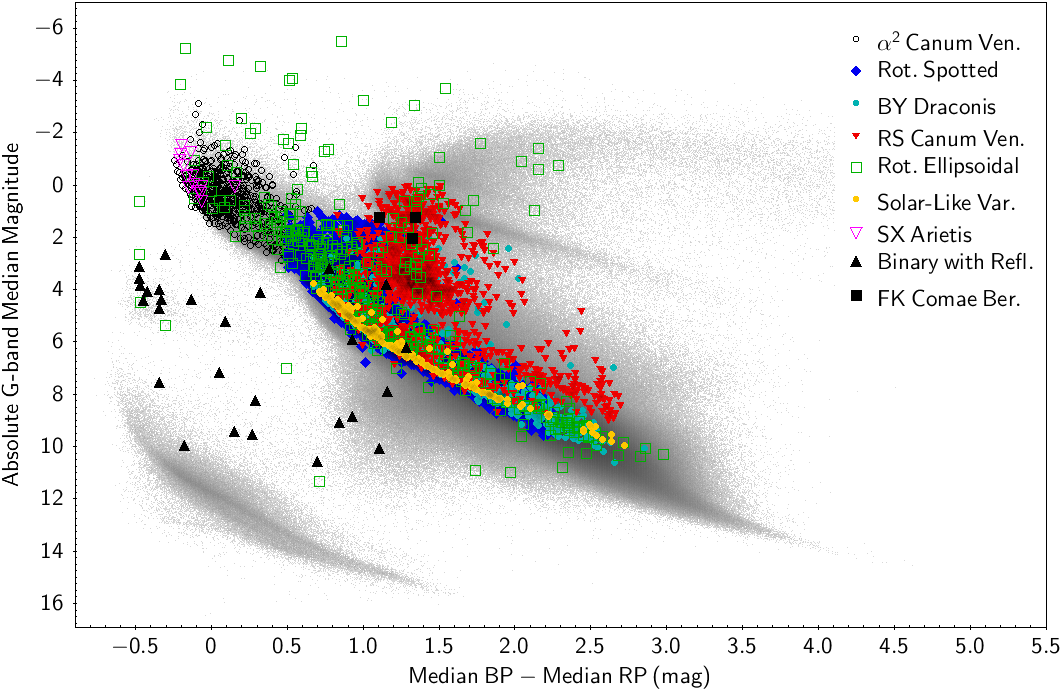}
   \caption{Same as Fig.~\ref{fig:varTypeCMD_Puls}, but for rotational-induced variability types.
   }
    \label{fig:varTypeCMD_Rot}
\vspace{0.7cm}
 \includegraphics[height=0.45\textheight]{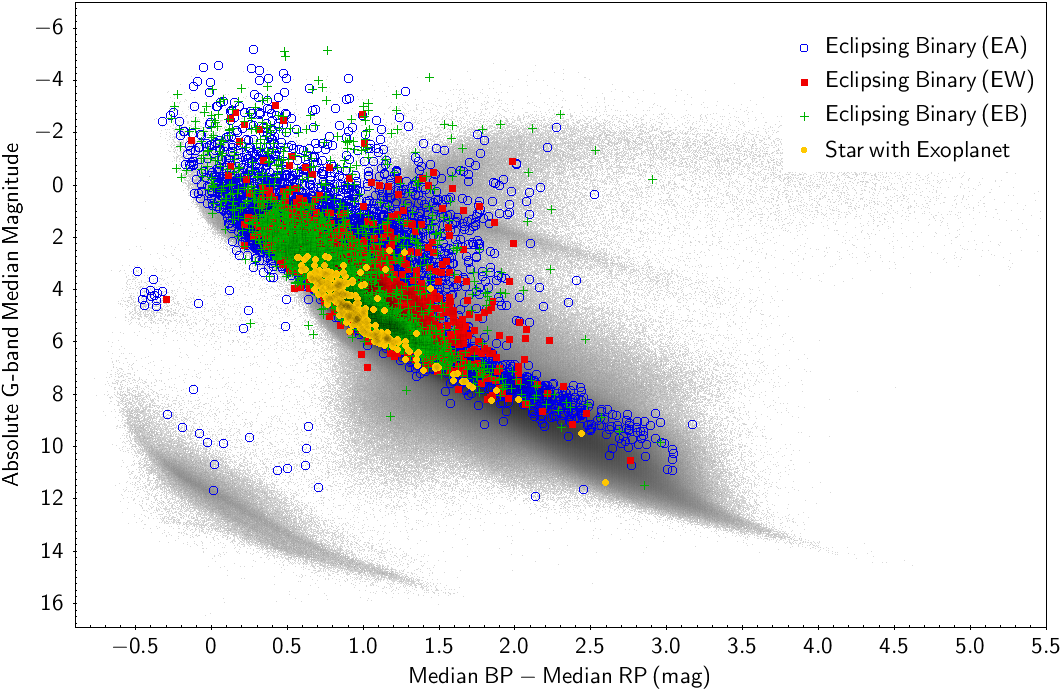}
   \caption{Same as Fig.~\ref{fig:varTypeCMD_Puls}, but for eclipsing binaries (of types EA, EB, and EW) and known host-stars that show exoplanet transits. As expected,  eclipsing binaries can be anywhere in the CaMD, which explains why they are the main source of contamination of pulsating stars, for instance.
   }
    \label{fig:varTypeCMD_Ecl}
\end{figure*}

Figure~\ref{fig:varTypeCMD_Rot} shows stars whose 
variability is induced by rotation. 
There are three primary categories: spotted stars,  stars deformed by tidal interactions, and objects whose variability is due to light reflected by a companion. Following the nomenclature in the literature (Table~\ref{tab:literature}), we list the following variability classes separately, although we note occasional overlaps among the definitions of these variability classes. The following types are included in Fig.~\ref{fig:varTypeCMD_Rot}:
\begin{enumerate}
        \item $\alpha^2$\,Canum Venaticorum stars, which are highly magnetic variable Bp and Ap MS stars. 
        \item Spotted stars, which show rotational modulation variability from spots.
        \item BY Draconis stars, which are main-sequence stars with late spectral types (K and M) that exhibit quasi-periodic light curves due to spots and chromospheric activity.
        \item RS Canum Venaticorum stars, which are spotted stars whose rotation-induced variability is frequently accompanied by other phenomena, such as eclipses and flares.
        \item Ellipsoidal variables, which show variability (without eclipses) due to orbital motion of a star that is distorted by a stellar companion.
        \item Solar-like stars with magnetic activity. Stars of this type in Fig.~\ref{fig:varTypeCMD_Rot} are limited to a catalogue focused on the Pleiades, which explains the thin distribution of the main sequence. We can see a hint of the binary sequence.
        \item SX\,Arietis stars, which are similar to $\alpha^2$\,Canum Venaticorum stars but have a higher temperature. We note that some overlap of the two distributions occurs for these two variability types.
        \item Binary systems with a strong reflection component in the light curve with re-radiation of the hotter star's light from the cooler companion's surface.
        \item FK Comae Berenices stars, which are spotted giant stars.
\end{enumerate}

Figure~\ref{fig:varTypeCMD_Rot} shows the following properties, among other things.
\begin{itemize}
    \item RS Canum Venaticorum stars are significantly brighter than BY Draconis stars near the bottom of the main sequence (at cool temperatures).
    \item The reflection binary class is primarily present among very compact (subdwarf) stars; there is a cluster near absolute mag 4, $\mbox{\gbp}-\mbox{\grp} \sim -0.4$ mag.
    \item There seems to be a dearth of rotational spotted variables around $\mbox{\gbp}-\mbox{\grp} \sim 0.4$, which corresponds with the transition region of stars with a radiative versus convective outer envelope.
    \item SX Arietis stars form a fairly well-defined hot-temperature envelope of the most luminous $alpha^2$ Canum Venaticorum variables.
\end{itemize}

Figure~\ref{fig:varTypeCMD_Ecl} shows eclipsing binary systems as well as stars that have been identified to host exoplanets through the transit method. Symbols distinguish the following sub-classes:
\begin{enumerate}
        \item Eclipsing binaries of type EA; the prototype is Algol. Binaries with spherical or slightly ellipsoidal components with well-separated, nearly constant light curves in between minima. Secondary minima can be absent.
        \item Eclipsing binaries of type EB; the prototype is $\beta$\,Lyrae. Binaries with continuously changing light curves and not clearly defined onsets or ends of eclipses. Secondary minima are always present, but can be significantly less deep than primary minima. 
        \item Eclipsing binaries of type EW; the prototype is W Ursae Majoris. The components are nearly or actually in contact and minima are virtually equally strong. Onsets and ends of minima are not well defined.
        \item Stars known to exhibit exoplanetary transits from the literature. 
\end{enumerate}

Based on Fig.~\ref{fig:varTypeCMD_Ecl}, we observe that
EA stars are present almost throughout the CaMD. There are
groups of EB stars that are overluminous compared to the white dwarf sequence, which are likely white dwarf stars with main-sequence companions. Moreover, the majority of the stars that host exoplanets are identified by the Kepler spacecraft, and only very few of them have detectable transits in the \gaia data because of the different photometric precision and time sampling.

\begin{figure*}
\centering
 \includegraphics[height=0.45\textheight]{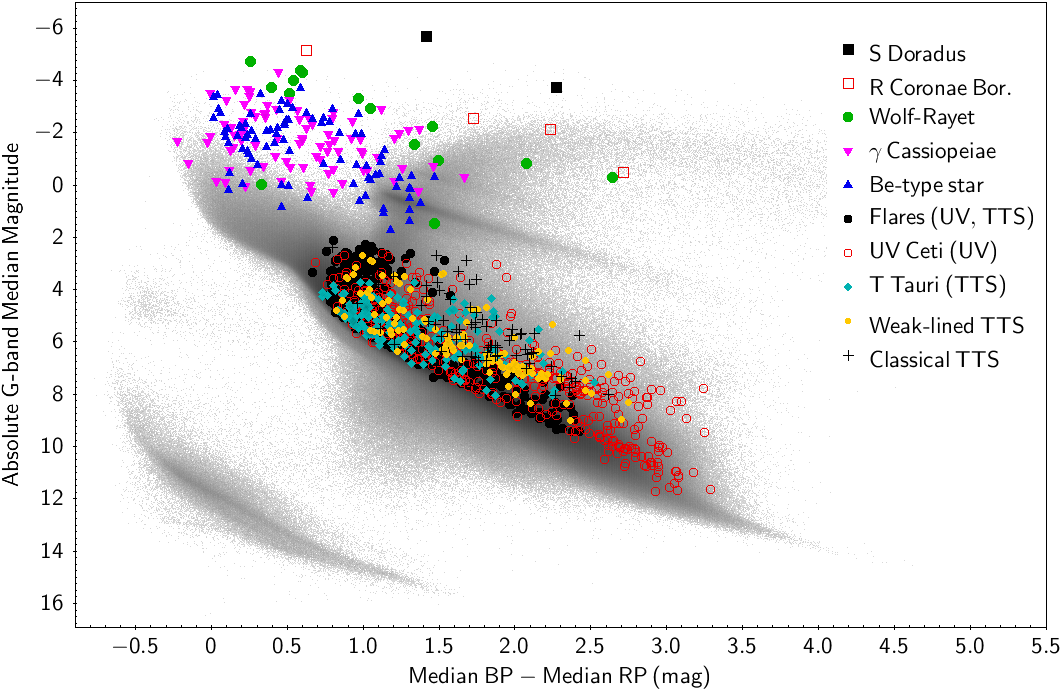}
   \caption{Same as Fig.~\ref{fig:varTypeCMD_Puls}, but for eruptive variability types.
 }
    \label{fig:varTypeCMD_Erup}
\vspace{0.7cm}
 \includegraphics[height=0.45\textheight]{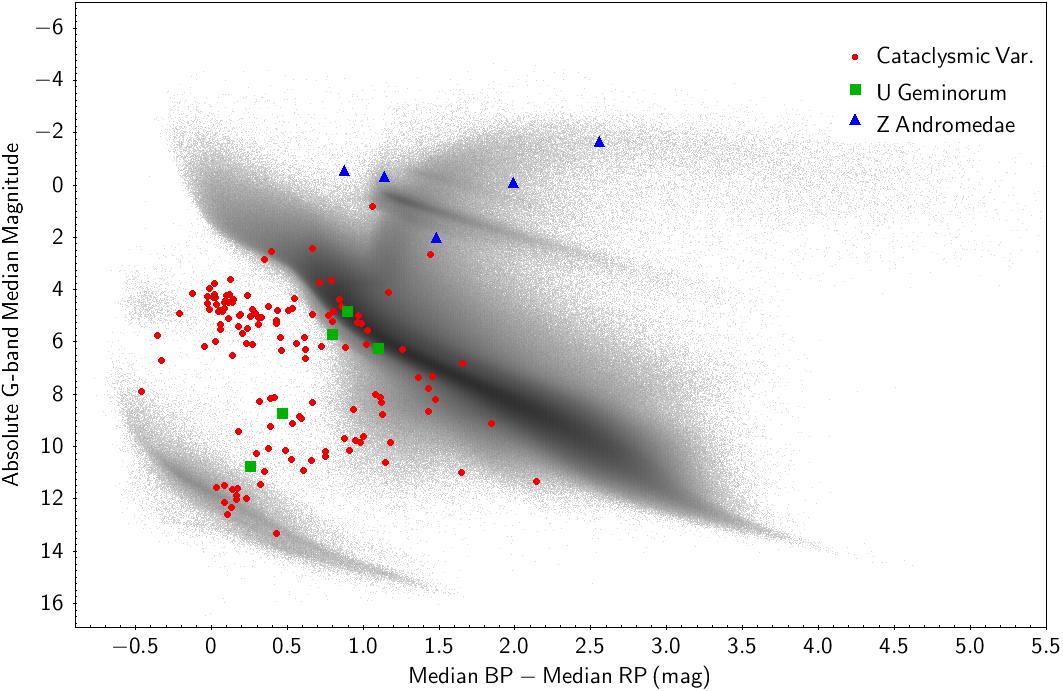}
   \caption{Same as Fig.~\ref{fig:varTypeCMD_Puls}, but for cataclysmic variables and some sub-types. 
}
    \label{fig:varTypeCMD_Catac}
\end{figure*}

\subsection{Eruptive and cataclysmic variables\label{sec:eruptiveCaMD}\label{sec:cataclysmicCaMD}}

Figure~\ref{fig:varTypeCMD_Erup} focuses on eruptive variable stars. As for the rotationally induced variables, we adopt the nomenclature from the literature (see Table~\ref{tab:literature}), which includes partially overlapping definitions. The following types are considered. 
\begin{enumerate}
        \item S Doradus stars, also known as luminous blue variables, are massive evolved stars that feature major and irregular photometric variations that are due to heavy mass loss by a radiation-driven wind.
        \item R Coronae Borealis stars, which are carbon-rich supergiants that emit obscuring material and as a consequence
    have drastic rapid dimming phases.
        \item Wolf-Rayet (WR) stars, which are the almost naked helium core that is left over from originally very high mass evolved stars. They feature strong emission lines of hydrogen, nitrogen, carbon, or oxygen. WR stars undergo very fast mass loss and can be significantly dust-attenuated.
        \item $\gamma$\,Cassiopeiae stars and stars with B spectral types that exhibit hydrogen emission lines, that is, Be stars. These are emitting shell stars. During their `eruptive' phenomena, they become brighter.
        \item Flare stars, which are magnetically active stars that display flares. This category incudes many subtypes of magnetically active stars, such as UV Ceti-type, RS CVn-type, and T Tauri stars.
        \item UV Ceti stars, which usually are K-M dwarfs that show flares.
        \item T Tauri stars (classical and weak lined), which are young pre-main sequence stars that either accrete strongly (classical) or show little sign of accretion (weak lined). These stars show variability that is due either to magnetic activity (e.g. rotational modulation and flares) or accretion (quasi-periodic, episodic, or stochastic variations), in addition to  pulsations that may also occur in some of them.
\end{enumerate}

In Fig.\,\ref{fig:varTypeCMD_Erup}, we notice the absence of eruptive variables among hot main-sequence stars (non-supergiants). This region is populated by pulsating stars, such as $\gamma$ Doradus and $\delta$ Scuti stars, cf. Fig.\,\ref{fig:varTypeCMD_Puls}. Moreover, WR stars, R Coronae Borealis stars, and S Doradus stars are among the most luminous stars in this diagram.

Figure~\ref{fig:varTypeCMD_Catac} illustrates three types of cataclysmic variables.
\begin{enumerate}
        \item Cataclysmic variables (generic class), typically novae and dwarf novae involving a white dwarf. Many of these stars are situated between the main and white dwarf sequences. 
        \item U Geminorum stars, which are dwarf novae that in principle consist of a white dwarf with a red dwarf companion that experiences mass transfer. 
        \item Z Andromedae stars, which are symbiotic binary stars composed of a giant and a white dwarf.
        \end{enumerate}
Further information on cataclysmic variables can be found for example in 
\citet{2003cvs..book.....W} and \citet{2001cvs..book.....H}.

In Fig.~\ref{fig:varTypeCMD_Catac}, we note a clump of cataclysmic variables located in the ZZ~Ceti variability strip near $G \sim 12$ and $\mbox{\gbp}-\mbox{\grp}$ $\sim 0.1$. 
The most significant clump of cataclysmic variables is near $G \sim 4$ and $\bpminrp \sim 0.1$ mag. These are probably binary systems with stars from the extreme horizontal branch and the main sequence.

\section{Variable object fractions in the CaMD
\label{sect:fraction} }
\begin{figure*}[t]
\centering
 \includegraphics[width=175mm]{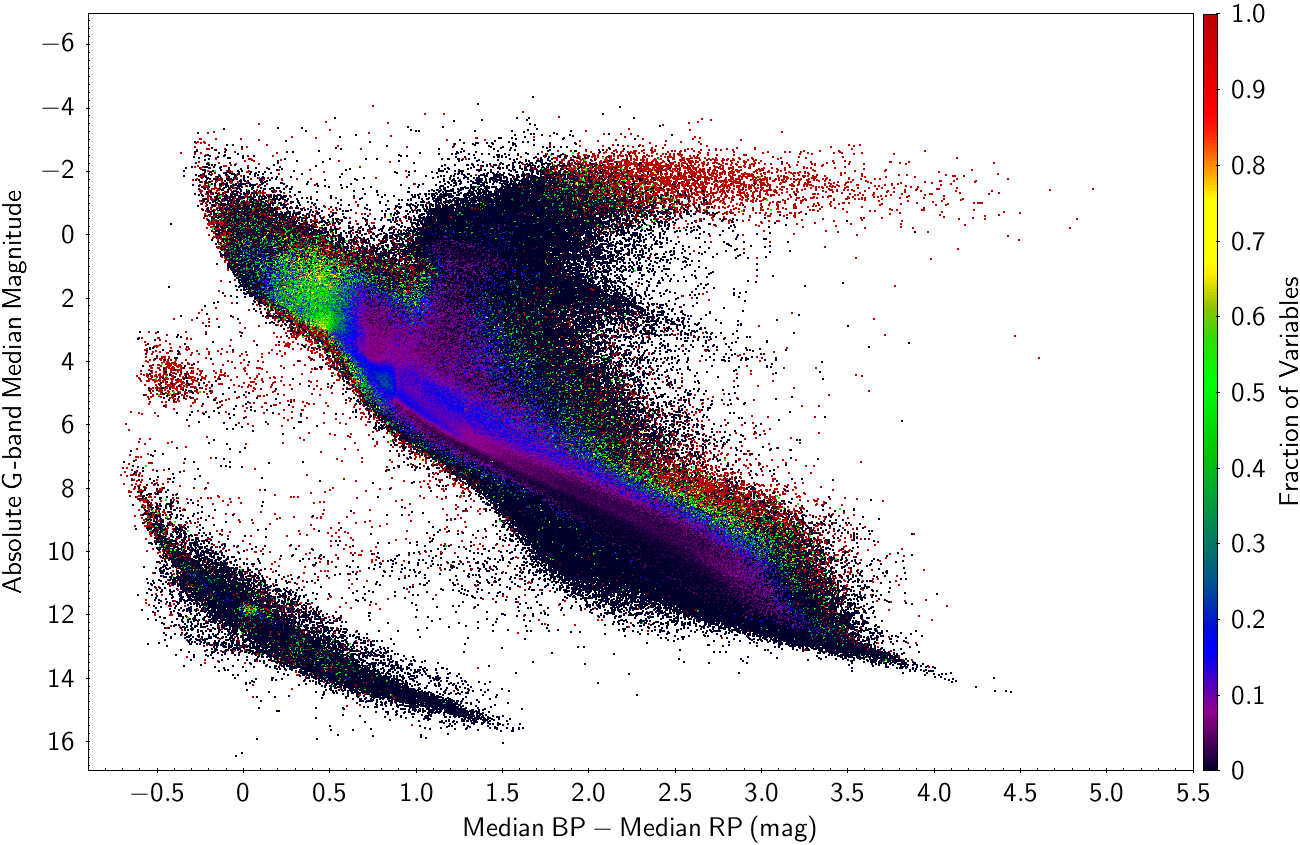}
   \caption{Variable object fraction 
   in the CaMD shown as a colour scale as labelled. This figure is not based on variable objects from the literature. Instead, variability is detected directly using \gaia data and employing supervised classification for sources with at least 20 observations in the \gmag, \gbp, and \grp bands.
   All objects satisfy the selection criteria described in Appendix~\ref{app:selection}, but with more restrictive constraints on the parallax precision (\texttt{parallax\_over\_error}~$>20$) and  on the parallax value ($\varpi > 1$\,mas), which limits the sample (size, extinction, and reddening). In order to reduce the extinction effect, objects at low Galactic latitudes (from $-$5 to 5~deg) are excluded. About 9\% of the 13.5~million stars that satisfy these criteria are variable. Some of the bins (especially the outlying ones) can contain only a few or even single sources. The condition on the relative precision of \gbp measurements introduces artificial cuts in the distributions of low-mass main-sequence stars and red (super)giants.
   } 
    \label{fig:FractionInHRD}
\end{figure*}

The different types of brightness variations as presented in the CaMD may strongly depend on the colour and absolute magnitude, as seen in Sect.~\ref{sect:location}, because they are driven by different physical mechanisms. Similarly, the variable object fraction, which is defined as the number of variable objects per colour-magnitude bin divided by the total number of objects in the same bin, is expected to depend on the location in the CaMD.  The variable object fraction was previously determined based on variable objects detected in the \textit{Hipparcos} time series \citep{HIPPARCOS_PERIODIC_ESA_1997}, for example. Here we significantly expand this investigation using 13.5 million stars with heliocentric distances of up to $1$\,kpc that satisfy the astrometric and photometric selection criteria listed in Appendix~\ref{app:selection} as well as (a) at least 20 observations in the \gmag, \gbp, and \grp bands, and (b) a relative parallax uncertainty of $< 5\%$~. In order to reduce the number of objects that are affected by significant extinction, stars at low Galactic latitudes (from $-$5 to 5~deg) are excluded. 
This effectively reduces the number of disc variables such as classical Cepheids and $\beta$\,Cephei stars.

Fig.~\ref{fig:FractionInHRD} illustrates this \gaia-based high-resolution map of the variable object fraction in the CaMD at the precision level of approximately 5--10~mmag.
Variability is identified in about 9\%~of the stars based on a supervised classification of \gaia sources. This method depends heavily on the selection of the training set of constant and variable objects. Minor colour-coded features 
can be due to training-set related biases. The detection of variability further depends upon the amplitude of the variables, their apparent magnitude distribution, and the instrumental precision. 
The accuracy of the fraction of variables is also affected by the number of sources per bin of absolute magnitude and colour, which can be as low as one in the tails of the two-dimensional source number density distribution.

Figure~\ref{fig:FractionInHRD} contains many informative features, despite possible biases.
Future data releases will significantly improve upon Fig.~\ref{fig:FractionInHRD} by correcting for reddening and extinction and using a larger number of objects with more accurate source classifications.
For the time being, we remark the following.
\begin{itemize}
  \item The classical instability strip is  clearly visible with a variability in about 50-60\%~of the stars (although extinction  limits the precision of this estimate).
  \item For evolved stars, red giants, and asymptotic giant branch stars, we find that higher luminosity and redder colour implies a higher probability of variability. 
  \item The red clump has a very low fraction of variable stars in the \gaia data. \textit{Kepler} photometry of red clump stars has revealed complex variability at the micro-mag level that has been used extensively for asteroseismology, cf. Sect.~\ref{sect:introduction} and references therein.
  \item The classical ZZ\,Ceti location is extremely concentrated in colour and magnitude, with variability in about half of the stars. The concentration is due to the partial ionisation of hydrogen in the outer envelope of white dwarfs, which is developed only in extremely narrow ranges of effective temperatures \citep[see][]{FontaineBrassard2008}.
  \item Extreme horizontal branch stars show a high probability of variability.
  \item The hottest and most luminous main-sequence stars are very frequently variable.
   \item There is a clear gradient towards larger fractions of variables above the low-mass main-sequence stars. 
\end{itemize}

\section{Variability amplitudes in the CaMD \label{sect:variationlevel} }
\begin{figure*}[t]
\centering
  \includegraphics[width=.95\textwidth]{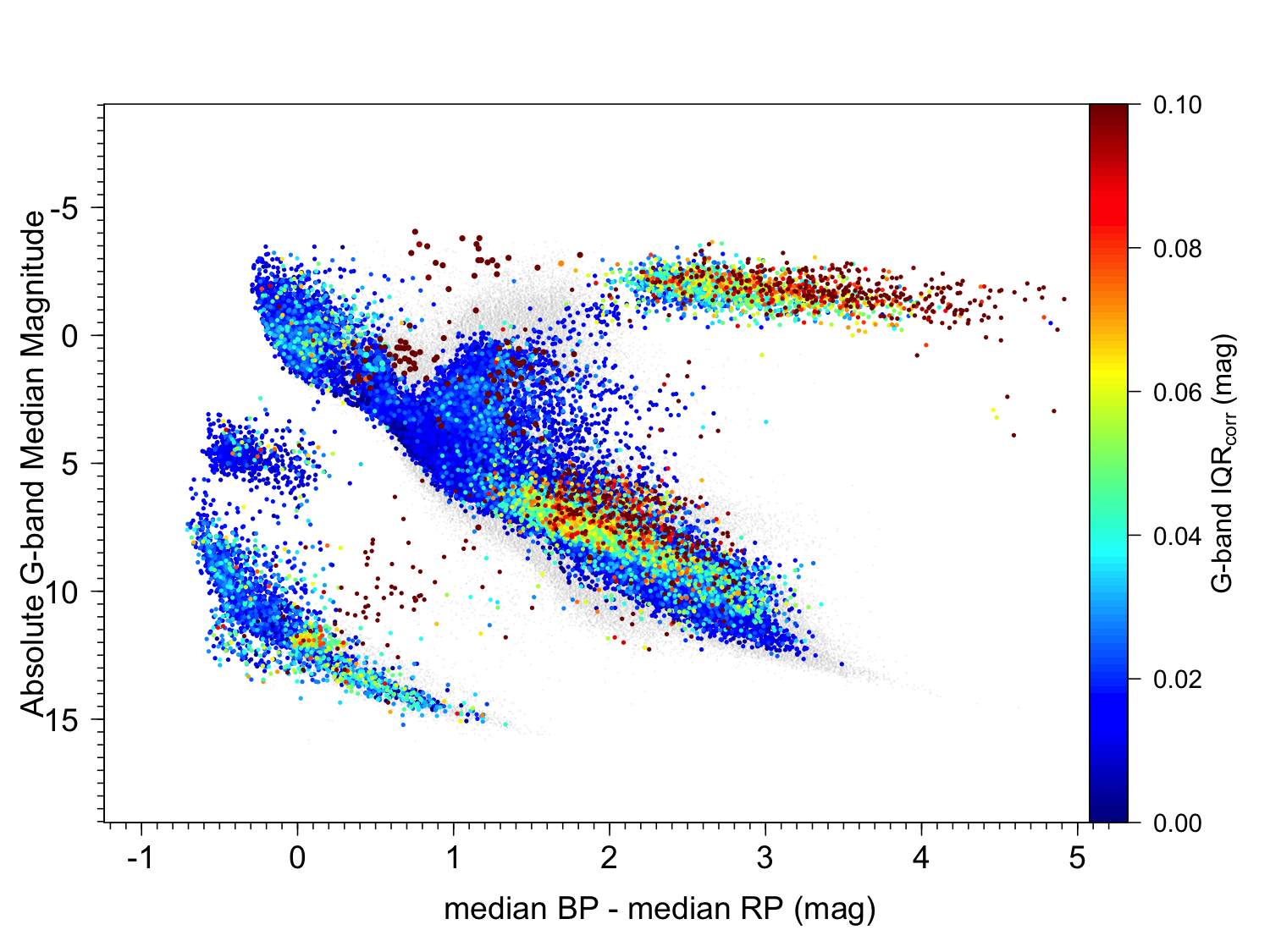}
 
   \caption{
   Amplitude of variability in the CaMD based on a selection of classified variables within 1~kpc and with a relative uncertainty for the parallax of 5\%. The colour scale shows the corrected \gmag-band IQR (see text) with a cut-off at 0.1~mag to emphasise the low- and mid-level variability. The background points in grey represent classified constant stars.
   All objects satisfy the selection criteria described in Appendix~\ref{app:selection}, in addition to the stricter conditions on parallax and its precision, as mentioned above.
   The effects of interstellar extinction are not corrected for.
   } 
    \label{fig:AmplitudeInHRD}
\end{figure*}

Figure~\ref{fig:AmplitudeInHRD} shows variability amplitudes as a function of position in the CaMD. Here, we quantify variability amplitudes 
using the \gmag-band inter-quartile range (IQR). Objects are selected according to the general criteria described in Appendix~\ref{app:selection}, with stricter conditions on the parallax (greater than 1~mas) and its relative precision (better than 5\%). 
To prevent the false impression that faint (and very bright) sources have intrinsically higher amplitudes, we corrected for the instrumental spread of the IQR as a function of the median \gmag magnitude. This correction was determined using sources that are classified as constant in the all-sky classification \citep{RimoldiniEtAl2018} and subtracted in quadrature from the measured IQR.
Instead of plotting individual data points in Fig.~\ref{fig:AmplitudeInHRD}, we show the (colour-coded) mean of the corrected \gmag-band IQR of sources within each square bin measuring 0.02~mag in both colour and magnitude after trimming the top and bottom 5\%.
This binning was applied to each variability type individually, and cuts were applied to select minimum classification probabilities per type to minimise incorrect classifications. We emphasise the location of variable object classes that feature large amplitudes by plotting classes with higher IQR on top of variability classes with lower IQR. 

\begin{figure}[h]
\centering
 \includegraphics[width=90mm]{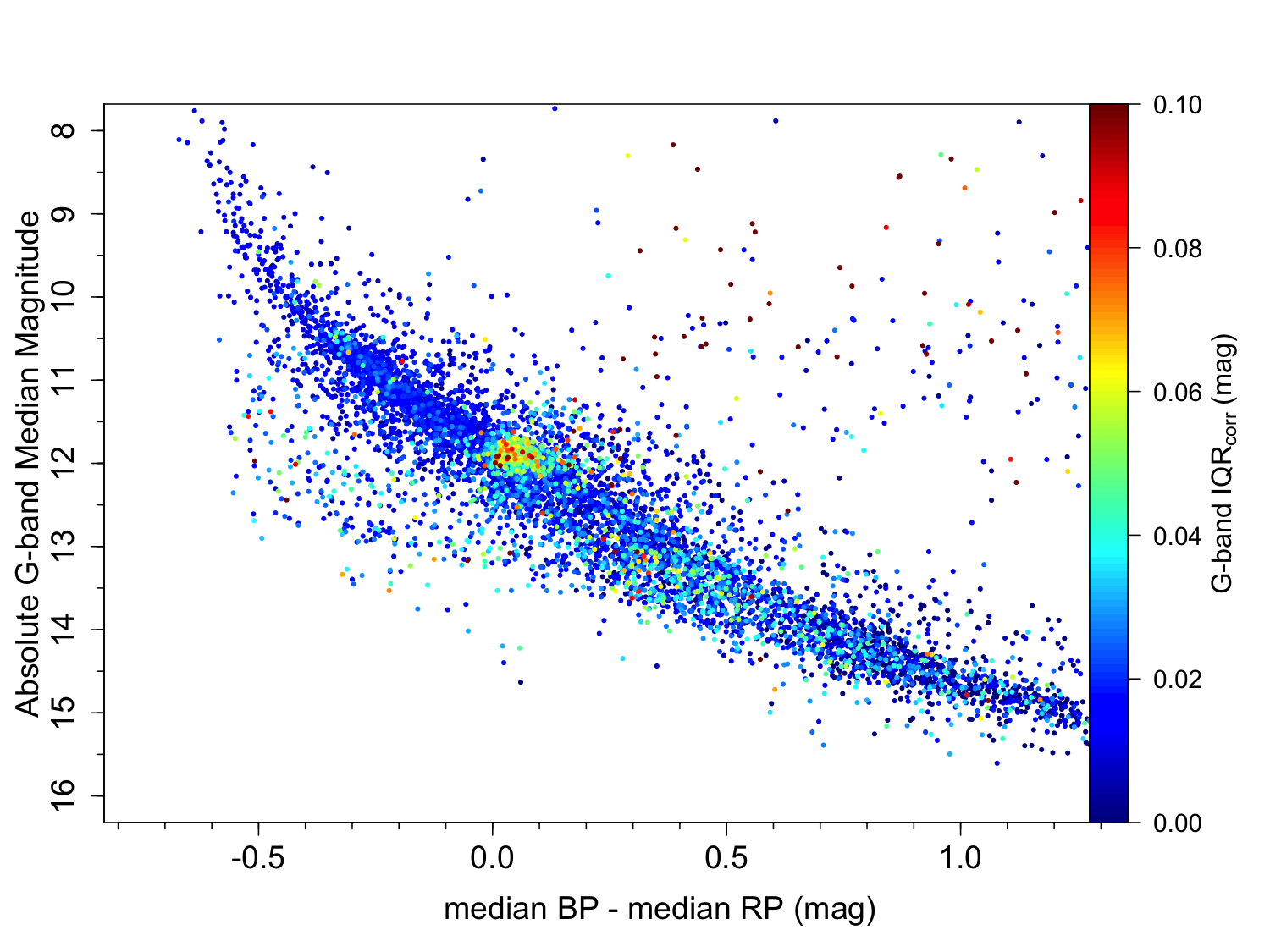}
   \caption{Same as Fig.~\ref{fig:AmplitudeInHRD}, but focusing on the white dwarf sequence and plotting all classified variables within 200~pc with a relative uncertainty for the parallax better than 5\%. A close inspection of this sequence reveals amplitudes at the level of 40 mmag in various regions.
   } 
    \label{fig:AmplitudeWDInHRD}
\end{figure}

Figure~\ref{fig:AmplitudeInHRD} contains the following stellar variability types based on the 
all-sky classification \citep{RimoldiniEtAl2018}:
$\alpha^2$\,Canum Venaticorum, $\alpha$\,Cygni, $\beta$\,Cephei, cataclysmic, classical Cepheids, $\delta$\,Scuti, $\gamma$\,Cassiopeiae, $\gamma$\,Doradus, Mira, ellipsoidal, RR\,Lyrae of Bailey's type ab and~c, semiregular, slowly pulsating B stars, solar-like variability due to magnetic activity (flares, spots, and rotational modulation), SX\,Arietis, and SX\,Phoenicis. We did not include other classes \citep[listed in][]{EyerGaiaDoc2018} for clarity or because there were too few objects.    
We note that any specific selection criteria applied to the objects shown in Fig.~\ref{fig:AmplitudeInHRD}  introduce biases that can highlight or diminish the prominence of certain phenomena. Nevertheless, Fig.~\ref{fig:AmplitudeInHRD} provides a first detailed illustration of some of the most important amplitude-related variability features in the CaMD. 
A number of clumps and instability regions are visible in Fig.~\ref{fig:AmplitudeInHRD}, which are related to the variability classes described in Sects.~\ref{sect:location} and~\ref{sect:fraction}. We notice the following features.
\begin{itemize}
    \item The classical instability strip that contains classical Cepheids and RR~Lyrae stars is not very prominent, although some clumps (in red or cyan) are apparent.
    \item The instability regions linked to SPB stars and $\beta$\,Cephei stars are  broad and uniform.
    \item Higher amplitude variations are clearly correlated with redder colours for long-period variables. 
    \item Variables with the highest amplitude (IQR $>0.1$ mag) occur in several regions in the CaMD, including the classical instability strip, long-period variables, below the red clump, above the main sequence of low-mass stars (in correspondence of the observed gradient in the fraction of variables), and between the white dwarf sequence and the main sequence.
    \item Significant amplitudes of $>0.04$ mag are found very frequently among the coolest white dwarfs.
    \item The stars between the main sequence and the white dwarfs sequence feature large variability amplitudes and extend into the clump of ZZ\,Ceti stars in the white dwarf sequence. This intermediate region is populated in particular by the high-amplitude cataclysmic variables, cf. Fig.\,\ref{fig:varTypeCMD_Catac}. 
    A closeup view of the white dwarf sequence is shown in Fig.~\ref{fig:AmplitudeWDInHRD}, which represents all classified variables within 200~pc. Each object is plotted without binning to emphasise the variability of the ZZ\,Ceti stars.
\end{itemize}

\section{Variability-induced motion in the CaMD \label{sect:motion} }
\begin{figure*}
\centering
 \includegraphics[width=175mm]{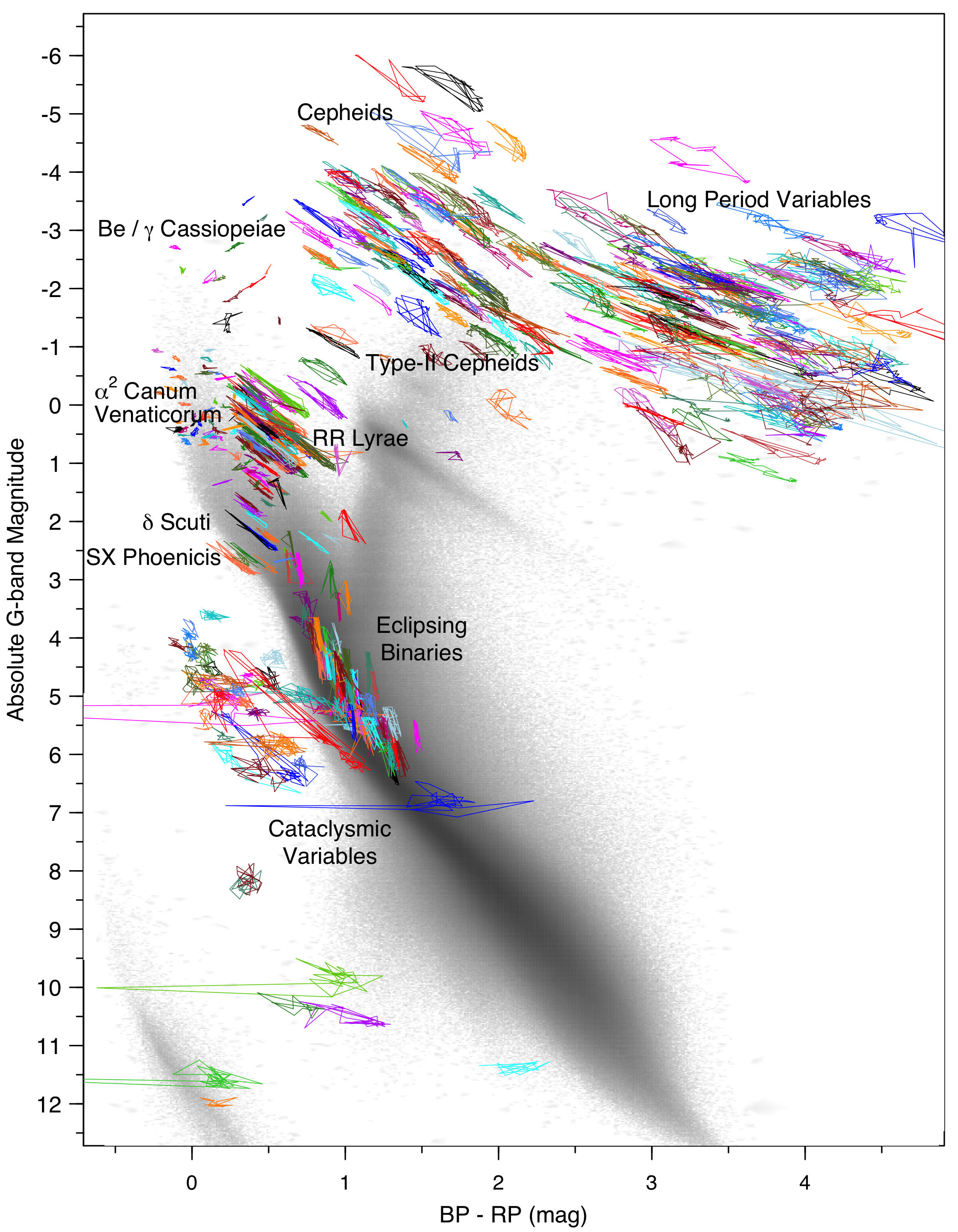}
 \caption{Motions of selected variable stars in the CaMD,
 highlighted by segments connecting successive absolute \gmag magnitudes and \bpminrp measurements in time with the same colour for the same source. Preferential directions and amplitudes of magnitude and colour
 variations can be inferred as a function of variability type ($\alpha^2$\,Canum
 Venaticorum, Be-type and $\gamma$\,Cassiopeiae, cataclysmic, classical and type-II~Cepheid, $\delta$\,Scuti and
 SX\,Phoenicis, eclipsing binary, long period, and RR\,Lyrae), as labelled
 in the figure. For clarity of visualisation, the selection of eclipsing binaries (and
 partially other types) was adjusted to minimise the overlap with other types. Selection criteria of all sources represented in colour or grey are the same as in Fig.~\ref{fig:varTypeCMD_Puls}. Additional conditions are described in the text.} 
 \label{fig:MotionInCMD}
\end{figure*}

In this section, we visualise the variability-induced motion of stars in the time-dependent CaMD using all-sky measurements made in the \gmag, \gbp, and \grp passbands. \gaia data are uniquely suited to create this time-dependent CaMD, since the different data types (astrometric and photometric in three bands) are acquired in a quasi-simultaneous fashion at many epochs that are distributed over a multi-year time span. 
The first of such representations, although much less detailed, was presented for individual classical Cepheids in the Milky Way \citep{1951ApJ...113..367E} and in Galactic star clusters \citep{1956PZ.....11..325K,1958ApJ...128..150S}. Similarly minded representations in the literature were based on data from the SDSS  \citep[mostly Galactic objects]{IvezicEtAl2003}, EROS \citep[LMC objects]{SpanoEtAl2009}, and, very recently, the \textit{HST} observations of M51 \citep{2018arXiv180405860C}. 

Figure~\ref{fig:MotionInCMD} illustrates the variability-induced motion of stars in the CaMD. As elsewhere in this paper, no correction for interstellar extinction is applied.
Individual stars are shown by differently (arbitrarily) coloured lines that connect successive absolute \gmag magnitudes and \bpminrp measurements, that is, the observations are ordered in time as opposed to variability phase. This choice was made to avoid uncertainties related to phase-folding the relatively sparsely sampled light curves based on 22 or fewer months of observations and to include both periodic and non-periodic variable objects.

Figure~\ref{fig:MotionInCMD} is limited to a subset of all available variable stars in order to avoid overcrowding the diagram. 
As a preview for future data releases, we include here the variability-induced motions of some stars whose time series and variability types are not published in DR2 (but which are available as online material).
Figure~\ref{fig:MotionInCMD} includes the following variability types as defined in Sec.\,\ref{sect:location}:
$\alpha^2$\,Canum Venaticorum variables, B-type emission line
/$\gamma$\,Cassiopeiae stars, cataclysmic variables, classical and type-II Cepheids, $\delta$\,Scuti stars, eclipsing binaries, 
RR~Lyrae stars, long-period variables, and SX\,Phoenicis stars.
All sources shown satisfy the general criteria described in Appendix~\ref{app:selection} and typically have at least ten available observations \footnote{The minimum number of observations per source is increased to 20 in the case of long-period variables, but the condition on the number of observations is removed for cataclysmic variables.}.
We further prioritised the selection of objects featuring 
wider ranges of variations in the \gmag band (with a minimum of about
0.1~mag)\footnote{A minimum range in the \gmag band is not required for $\alpha^2$\,Canum Venaticorum stars and cataclysmic variables as their variability may be small in the `white' \gmag band.}. The number of  sources shown for each variability type ranges
from a few to several tens and was selected to ensure clarity in case of high source density or overlapping variability classes in certain regions of the CaMD. 
In order to limit the effect of outlying values, time-series data are filtered by operators as described in \citet{DPACP-49}, and the 10~\%~
of the brightest and faintest observations in the \grp band are excluded for sources with \bpminrp lower than 1.5~mag.
Non-variable objects are shown as a grey background to provide a visual reference for the variable object locations in the CaMD. 
These stars satisfy the criteria described in Appendix~\ref{app:selection} as well as the stricter condition of $\varpi > 1$\,mas.
Stars whose variability is caused by different physical effects exhibit different motions within the time-dependent CaMD. We briefly summarise the different motions seen in Fig.\ref{fig:MotionInCMD} as follows.
\begin{enumerate}
        \item Pulsating stars, including long-period variables,
        Cepheids, RR Lyrae, and $\delta$\,Scuti/SX\,Phoenicis stars, exhibit a similar behaviour. These stars are bluer when brighter in \gmag, which illustrates that brightness variations of pulsating stars are dominated by the effect of change in temperature  rather than radius.
    For the longest-period variables, the 22-month time span of the \gaia data is similar to the pulsation cycle, so that in some cases, loop-like shapes are apparent.
        For variable stars with shorter periods (e.g. RR\,Lyrae stars or classical Cepheids), 
        successive measurements in time
        are not generally ordered in phase,
        so that an overall `envelope' of many cycles is revealed.
        \item The motions of eclipsing binary systems in the CaMD depends on the colour difference between the two stars. The components of eclipsing systems of the EW type have similar mass (and colour), leading to a rather vertically aligned motion (parallel to the absolute magnitude axis). For eclipsing binary systems with stars of different mass (and thus colour) close to the main sequence, the deepest eclipse is usually slightly redder, 
        since the secondary (less massive and redder) component eclipses part of the light of the primary star.
        The slope of the motion of eclipsing binaries in the CaMD is very different (much steeper) than the one of pulsating stars.
        \item Rotationally induced variables (shown here: $\alpha^2$\,Canum Venaticorum stars) feature small amplitudes in absolute \gmag and are rather horizontal in the CaMD.
        This is as expected from starspots, which have a lower temperature than the surroundings and hence absorb the light at bluer frequencies to re-emit it by back-warming effect at redder frequencies. Therefore, the magnitude change in a broad band like \gmag is smaller than it would be if measured in narrow bands.
        \item Eruptive stars (shown here: $\gamma$\,Cassiopeiae and Be-type stars) become redder when brighter because of additional extinction during their eruptive phase. The slopes of their motions in the CaMD therefore have the opposite sign with respect to the sign of pulsating stars.
        \item The variability of cataclysmic variables (shown here: novae) primarily features strong outbursts in the ultraviolet and blue part of the spectrum that are understood to be caused by mass transfer from donor stars in binary systems. These outbursts very significantly change the colour of the system towards bluer values.
\end{enumerate}

The current version of Fig.~\ref{fig:MotionInCMD} represents a first step towards a more global description of stellar variability.
The motions described by the variable stars in the time-dependent CaMD provide new perspectives on the data that can be exploited as variable star classification attributes to appreciably improve the classification results. 
\gaia data will definitively help identify misclassifications
and problems in published catalogues, thanks to its astrometry and the quasi-simultaneous measurements.

In future \gaia data releases, there will be more data points per source, which will enable us to refine Fig.~\ref{fig:MotionInCMD}.
In particular, periods can be determined with an accuracy inversely proportional to the total \gaia time base for periodic objects. In this way, the motion in the CaMD can be represented more precisely by connecting points that are sorted in phase (rather than in time). this leads to Lissajous-type configurations for pulsators. 
For sufficiently bright stars, radial velocity time series will add a third and unprecedented dimension to Fig.~\ref{fig:MotionInCMD}. 

An animated version of Fig.~\ref{fig:MotionInCMD} is provided at \url{https://www.cosmos.esa.int/web/gaia/gaiadr2_cu7}. We provide online material that includes the time series in the G, BP, and RP  bands of the selected field-of-view transits  for 224 sources that are not published in Gaia DR2, but are plotted in Fig.~\ref{fig:MotionInCMD}.

\section{Conclusions \label{sect:conclusions} }
The \gaia mission enables a comprehensive description of phenomena related to stellar variability. We here focused on stellar variability in the CaMD and showed locations that are occupied by different variability types as well as variable object fractions, variability amplitudes, and variability-induced motions that are described by different variability classes in the CaMD. 

The wealth of information related to variable stars that is contained in \gaia DR2 is unprecedented for the Milky Way. 
The CaMD can provide guidance for further detailed studies, which can focus on individual regions or clumps, for instance, to investigate the purity of instability strips and how sharply such regions are truly defined or how they depend on chemical composition. Of course, additional work is required to this end, and accurately correcting for reddening and extinction will be crucial. 
The (time-dependent) CaMD will play an important role for improving the variable star classification by providing additional attributes, such as the expected direction of variability for specific variable classes, and for illustrating stellar variability to non-expert audiences.

The CaMD of variable stars can further point out interrelations between variability phenomena that are otherwise not easily recognised, and it might be able to identify new types of variability. Detailed follow-up observations from the ground will help correct previous misclassifications and enable in-depth studies of peculiar and particularly interesting objects. Based on the variable stars that reside in the Milky Way, as presented here, it will be possible to obtain data with particularly high S/N, for example through high-resolution spectroscopy. Finally, the  observed properties of variable stars in the CaMD, such as instability strip boundaries or period-luminosity relations, provide crucial input and constraints for models describing pulsational instability, convection, and stellar structure in general.

Future \gaia data releases will further surpass the variability content of this second data release\footnote{cf. \url{https://www.cosmos.esa.int/web/gaia/release}}.
By the end of mission, \gaia data are expected to comprise many tens of millions of variable celestial objects, including many additional variability types, as well as time series BP and RP spectra. Eventually, time series of radial velocities and spectra from the radial velocity spectrometer will be published for subsets of variables. Finally, the variability classification of future \gaia data will also make use of unsupervised clustering techniques aimed at discovering entirely new (sub-)clusters and classes of variable phenomena.

\begin{acknowledgements}
We would like to thank Laurent Rohrbasser for tests done on the representation of time
series.
This work has made use of data from the ESA space mission {\it Gaia}, processed by the
{\it Gaia} Data Processing and Analysis Consortium (DPAC). Funding for the DPAC has been
provided by national institutions, some of which participate in the {\it Gaia}
Multilateral Agreement, which include, for Switzerland, the Swiss State Secretariat for
Education, Research and Innovation through the ESA Prodex program, the ``Mesures
d’accompagnement”, the ``Activit\'{e}s Nationales Compl\'{e}mentaires”, the Swiss National
Science Foundation, and the Early Postdoc.Mobility fellowship; for Belgium, the BELgian
federal Science Policy Office (BELSPO) through PRODEX grants; for Italy, Istituto
Nazionale di Astrofisica (INAF) and the Agenzia Spaziale Italiana (ASI) through grants
I/037/08/0,  I/058/10/0,  2014-025-R.0, and 2014-025-R.1.2015 to INAF (PI M.G. Lattanzi);
for France, the Centre National d'Etudes Spatiales (CNES). Part of this research has received funding from the European Research Council (ERC)
  under the European Union's Horizon 2020 research and innovation programme
  (Advanced Grant agreements N$^\circ$670519: MAMSIE ``Mixing and Angular
  Momentum tranSport in MassIvE stars'').\\
This research has made use of NASA's Astrophysics Data System, the VizieR catalogue access tool, CDS, Strasbourg, France, and the International Variable Star Index (VSX) database, operated at AAVSO, Cambridge, Massachusetts, USA.\\
We gratefully acknowledge Mark Taylor for creating the astronomy-oriented data handling and visualization software TOPCAT
\citep{2005ASPC..347...29T}.

\end{acknowledgements}

\begin{appendix}

\section{Literature for each variability type \label{app:literature}}
See Table~\ref{tab:literature} for details on the references from the literature regarding the
objects included in Figs.~\ref{fig:varTypeCMD_Puls}--\ref{fig:varTypeCMD_Catac} and \ref{fig:MotionInCMD}.

\begin{table*}
\caption{Literature references of stars as a function of variability type and the corresponding number of sources depicted in Figs.~\ref{fig:varTypeCMD_Puls}--\ref{fig:varTypeCMD_Catac}, after selections based on reliability, photometric accuracy, and astrometric parameters (Appendix~\ref{app:selection}). Figure~\ref{fig:MotionInCMD} includes only subsets of variability types and of sources per type.} 
\label{tab:literature}     
\centering                         
\begin{tabular}{lllr}       
\hline\hline               
\noalign{\smallskip}
  Variability  & Type & Reference & \# Sources \\
\noalign{\smallskip}
\hline
\noalign{\smallskip}
Pulsating & $\alpha$ Cygni & Hip97, VSX16 & 17\\
 & $\beta$ Cephei & PDC05 & 20\\
 & Cepheid & ASA09, Hip97, INT12 & 155\\
 & $\delta$ Scuti & ASA09, Hip97, JD07, Kep11b, Kep11c, SDS12 & 724\\
 & $\gamma$ Doradus & FKA16, Kep11b, Kep11c, VSX16 & 561\\
 & Long Period Variable & ASA12, Hip97, INT12, Kep11b, NSV04 & 5221\\
 & PV Telescopii & VSX16  & 3\\
 & Rapidly Oscillating Am star & VSX16 & 8\\
 & Rapidly Oscillating Ap star & VSX16 & 25\\
 & RR Lyrae, fundamental mode (RRab) & ASA09, ASA12, Cat13a, Cat13b, Cat14b, Cat15, Hip97,  & 1676\\
  &            & INT12, LIN13, NSV06, VFB16, VSX16 \\
 & RR Lyrae, first overtone (RRc) & ASA09, ASA12, Cat13b, Cat14b, Hip97, INT12, Kep11b, & 611\\
 &          & LIN13, MA14, VFB16, VSX16 \\
 & RV Tauri & ASA12, Hip97, VSX16 & 48\\
 & Slowly Pulsating B star & IUE03, Hip97, PDC05 & 78\\
 & SX Phoenicis & ASA12, Hip97, VSX16 & 41\\
 & Type-II Cepheid & ASA12, Cat14b, Hip97, VSX16 & 21\\
 & V361 Hya (also EC 14026) & VSX16 & 41\\
 & V1093 Her (also PG 1716) & VSX16 & 1\\
 & ZZ Ceti & VSX16 & 61\\
\noalign{\smallskip}
\hline
\noalign{\smallskip}
Rotational & $\alpha^2$\,Canum Venaticorum & Hip97, VSX16 & 598\\
 & Binary with Reflection & VSX16 & 27\\
 & BY Draconis & VSX16 & 713\\
 & Ellipsoidal & ASA12, Cat14b, Hip97, Kep11b, VSX16 & 398\\
 & FK Comae Berenices & Hip97 & 3\\
 & Rotating Spotted & Kep15b & 16\,593\\
 & RS Canum Venaticorum & ASA12, Cat14b, Hip97, VSX16 & 1381\\
 & Solar-Like Variations & HAT10 & 176\\
 & SX Arietis & Hip97, VSX16 & 14\\
\noalign{\smallskip}
\hline
\noalign{\smallskip}
Eclipsing & EA, $\beta$~Persei (Algol) & ASA09, Cat14b, Hip97, LIN13, VSX16 & 8123\\
 & EB, $\beta$~Lyrae & ASA09, Cat14b, Hip97, LIN13, VSX16 & 3096\\
 & EW, W Ursae Majoris & ASA09, Hip97, VSX16 & 3248\\
 & Exoplanet & JS15 & 278\\
\noalign{\smallskip}
\hline
\noalign{\smallskip}
Eruptive & B-type emission-line star & ASA12, VSX16 & 86\\
 & Classical T Tauri Star & VSX16 & 75\\
 & Flares (UV, BY, TTS) & Kep11a, Kep13, Kep15a, MMT15 & 478\\
 & $\gamma$ Cassiopeiae & Hip97, VSX16 & 84\\
 & R Coronae Borealis & VSX16 & 4\\
 & S Doradus & ASA12, INT12 & 2\\
 & T Tauri Star (TTS) & VSX16 & 173\\
 & UV Ceti & INT12, VSX16 & 425\\
 & Weak-lined T Tauri Star & VSX16 & 119\\
 & Wolf-Rayet & INT12, VSX16 & 15\\
\noalign{\smallskip}
\hline
\noalign{\smallskip}
Cataclysmic & Cataclysmic Variable (generic) & Cat14a, OGL15, VSX16 & 132\\
 & U Geminorum & INT12, VSX16 & 4\\
 & Z Andromedae & INT12, VSX16 & 5\\
\noalign{\smallskip}
\hline                                  
\end{tabular}
\tablefoot{ASA09: \citet{ASAS_VARIABLES_KEPLER_PIGULSKI_2009};
ASA12: \citet{ASAS_VARIABLES_RICHARDS_2012}; 
Cat13a: \citet{CATALINA_RRAB_PAPER1_DRAKE_2013}; 
Cat13b: \citet{CATALINA_RR_PAPER2_DRAKE_2013}; 
Cat14a: \citet{CATALINA_CATACLYSMIC_VARIABLES_DRAKE_2014}; 
Cat14b: \citet{CATALINA_PERIODIC_VARIABLES_DRAKE_2014}; 
Cat15: \citet{CATALINA_RRAB_SSS_TORREALBA_2015}; 
FKA16: \citet{GDOR_ALICAVUS_2016}; 
HAT10: \citet{PLEIADES_HARTMAN_2010}; 
Hip97: \citet{HIPPARCOS_PERIODIC_ESA_1997}; 
INT12: \citet{INTEGRAL_OMC_VARIABLES_ALFONSO_GARZON_2012}; 
IUE03: \citet{SPB_NIEMCZURA_2003}; 
JD07: \citet{DSCT_GDOR_DEBOSSCHER_2007}; 
JS15: J. Southworth, \url{http://www.astro.keele.ac.uk/jkt/tepcat/observables.html} (as of Aug. 2015);
Kep11a: \citet{KEPLER_FLARES_2}; 
Kep11b: \citet{KEPLER_VARIABLES_DEBOSSCHER_2011}; 
Kep11c: \citet{DSCT_GDOR_UYTTERHOEVEN_2011}; 
Kep13: \citet{KEPLER_FLARES_1}; 
Kep15a: \citet{KEPLER_FLARES_3}; 
Kep15b: \citet{KEPLER_ROTATIONAL_MODULATION_REINHOLD_2015}; 
LIN13: \citet{LINEAR_PERIODIC_PALAVERSA_2013}; 
MA14: \citet{SDSS_PS1_CATALINA_RRL_ABBAS_2014}; 
MMT15: \cite{M37_FLARES}; 
NSV04: \citet{NSVS_RED_VARIABLES_WILLIAMS_2004}; 
NSV06: \citet{NSVS_RRAB_KINEMUCHI_2006}; 
OGL15: \citet{OGLE4_CATACLYSMIC_VARIABLES_20160718}; 
PDC05: P. De Cat, \url{http://www.ster.kuleuven.ac.be/~peter/Bstars/} (as of Jan. 2005); 
SDS12: \citet{SDSS_STRIPE82_DSCT_RR_SUVEGES_2012}; 
VFB16: \citet{RRL_IN_GLOBULAR_CLUSTER_OMEGACENTAURI}; 
VSX16: \citet{AAVSO_VSX_20160704, AAVSO_CDS, 2006SASS...25...47W}.} 
\end{table*}

\section{Selection criteria \label{app:selection}}
Astrometric and photometric conditions are applied to all CaMDs for improved accuracy of the star locations in such diagrams. 
Astrometric constraints include limits on the number of visibility periods (observation groups separated from other groups by at least four days) per source used in the secondary astrometric solution \citep{DPACP-31}, the excess astrometric noise of the source postulated to explain the scatter of residuals in the astrometric solution for that source \citep{DPACP-31}, and the relative parallax precision (herein set to 20\% although in other applications it was set to 5 or 10\%):
\begin{enumerate}
  \item \texttt{visibility\_periods\_used} $>$ 5; 
  \item \texttt{astrometric\_excess\_noise} $<$ 0.5~mas; 
  \item \texttt{parallax} $>$ 0~mas; 
  \item \texttt{parallax\_over\_error} $>5$.
\end{enumerate}
Photometric conditions set limits for each source on the relative precisions of the mean fluxes in the \gbp, \grp, and \gmag bands, as well as on the mean flux excess in the \gbp and \grp bands with respect to the \gmag band as a function of colour \citep{DPACP-40}:
\begin{enumerate}
  \setcounter{enumi}{4}
  \item \texttt{phot\_bp\_mean\_flux\_error}\,/\,\texttt{phot\_bp\_mean\_flux}~$<$~0.05;
  \item  \texttt{phot\_rp\_mean\_flux\_error}\,/\,\texttt{phot\_rp\_mean\_flux}~$<$~0.05;
  \item  \texttt{phot\_g\_mean\_flux\_error}\,/\,\texttt{phot\_g\_mean\_flux}~$<$~0.02;
  \item (\texttt{phot\_bp\_mean\_flux} + \texttt{phot\_rp\_mean\_flux}) / \\
   \{\texttt{phot\_g\_mean\_flux} * [1.2 + 0.03 * (\texttt{phot\_bp\_mean\_mag} - \texttt{phot\_rp\_mean\_mag})$^2$]\}  $<$ 1.2.
\end{enumerate}

The ADQL query to select a sample of sources that satisfy all of the above listed criteria follows.

\begin{verbatim}
SELECT TOP 10 source_id
FROM gaiadr2.gaia_source 
WHERE visibility_periods_used > 5
  AND astrometric_excess_noise < 0.5
  AND parallax > 0
  AND parallax_over_error > 5
  AND phot_bp_mean_flux_over_error > 20
  AND phot_rp_mean_flux_over_error > 20
  AND phot_g_mean_flux_over_error > 50
  AND phot_bp_rp_excess_factor < 1.2*(1.2+0.03*
      power(phot_bp_mean_mag-phot_rp_mean_mag,2))
\end{verbatim}

\end{appendix}

\bibliographystyle{aa}
\raggedbottom
\bibliography{local}

\end{document}